\documentclass[aps,prl,reprint,amsmath,amssymb,longbibliography,nofootinbib,floatfix]{revtex4-2}
\usepackage[T1]{fontenc}
\usepackage{bm,booktabs,graphicx}
\usepackage[colorlinks=true,linkcolor=blue,citecolor=blue,urlcolor=blue]{hyperref}
\graphicspath{{./}}
\newcommand{\E}{\mathbb E}
\newcommand{\diag}{\operatorname{diag}}
\newcommand{\cT}{\mathcal T}
\begin{document}
\title{Perfect resonance and fragile localization suppression in correlated disordered chains}

\author{Zhixuan Zhao}
\email{zhaozhixuan@stumail.ysu.edu.cn}
\author{Jun Li}
\email{ljcj007@ysu.edu.cn}
\affiliation{
	State Key Laboratory of Metastable Materials Science and Technology, Hebei Key Laboratory of Microstructural Material Physics, School of Science, Yanshan University, Qinhuangdao, 066004, China.}

\date{\today}
\begin{abstract}
Spatial correlations can suppress scattering in disordered chains and produce perfectly transmitting resonances. The practical value of this protection, however, depends not on the resonance peak itself but on the width of the surrounding transmission window and its sensitivity to local ordering errors. We show that a resonance can remain exactly transparent at its center while arbitrarily rare adjacent-swap errors restore an inverse localization length proportional to the square of the energy detuning. For lossless single-channel chains assembled from independent blocks of fixed length and composition, this positive quadratic term is guaranteed by a local scattering invariant and holds for every arrangement and every fixed swap probability between zero and one. With exact tuning and matched contacts, the central transmission remains unity. A microscopic quantum chain exhibits both higher-order suppression of scattering in the ideal recursive arrangement and the predicted response to local exchanges. These results reveal a limitation of spatial ordering that is invisible to a measurement at the resonance alone: spatial ordering protects the resonance peak, not the transport around it. The effect can therefore be tested experimentally by measuring transmission spectra before and after exchanges, without identifying microscopic defects.
\end{abstract}
\maketitle

Interference can prevent waves from propagating through a disordered
medium~\cite{Anderson1958}. Spatial correlations offer a way to control this
localization: a suitable sequence of scatterers can cancel reflection at
selected energies. The random-dimer and random-polymer models established
this mechanism~\cite{Dunlap1990,Jitomirskaya2003}, and microwave and optical
experiments have demonstrated that deliberately correlated disorder changes
wave transport~\cite{Kuhl2000,Naether2013}. For a finite device, the relevant
question is how much of the transmission spectrum remains useful when the
prescribed correlations are imperfect.

A transmitting resonance is a single-energy property. Its neighborhood
can be improved by making the localization rate vanish more rapidly as the
incident energy approaches resonance. Merging resonances in a random trimer
chain changes the characteristic window from $N^{-1/2}$ to
$N^{-1/4}$ for a chain of $N$ cells~\cite{Giri1993}. Related recursive
sequences suppress successive scattering orders, as in random multipolar
driving~\cite{Zhao2021,Mori2021} and its extension to transfer
matrices~\cite{Mo2025}. The physical benefit is weaker attenuation over a
larger energy interval at a given length. Its robustness depends on the
local errors introduced when the sequence is realized.

We consider errors that exchange two neighboring cells without changing
the number of cells of each type. In the chains studied here, these numbers
fix the resonance, whereas their spatial order determines nearby scattering.
An exchange can therefore preserve the peak transmission exactly while
changing the surrounding spectrum. Defects are known to restore localization
in correlated chains~\cite{Xiong1996}, and higher-order scattering can defeat
leading-order transport predictions based on the disorder spectrum~\cite{Dikopoltsev2022}.
The question here is whether one can choose a better arrangement that suppresses
local exchange errors while retaining the resonant condition.

We prove that every arrangement within a specified single-channel class
has an unavoidable response to randomly placed neighboring exchanges.
A local measure of their scattering strength is independent of where the
pair is embedded, and every allowed block must contain a pair that scatters at first order in energy.
Consequently, a nonzero density of such errors restores quadratic localization,
even though each complete sample remains perfectly transmitting at resonance.
A microscopic quantum chain connects this bound to high-order ideal designs
and finite-device transmission spectra. The resulting distinction suggests a direct experimental test: keep the cell parameters, counts, and contacts
fixed, exchange neighboring cells, and measure the spectrum around the
unchanged central peak.

\begin{figure*}[t]
 \includegraphics[width=\textwidth]{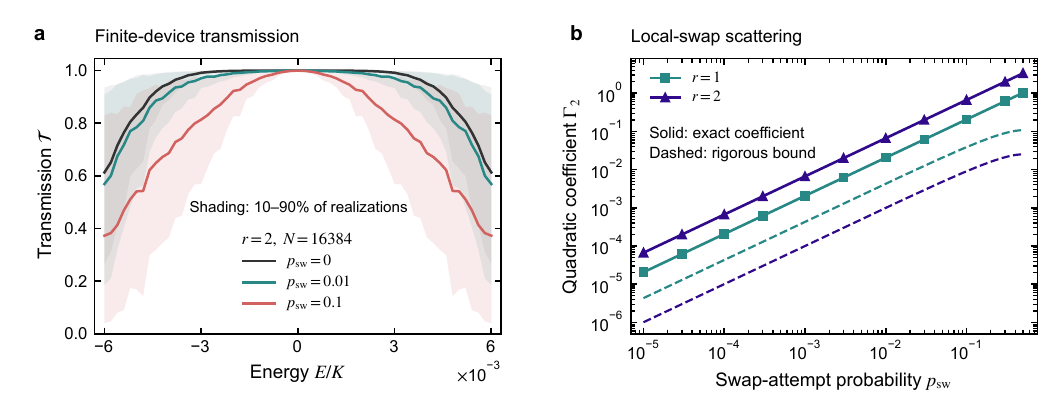}
 \caption{Local exchanges preserve the resonance but alter its neighborhood.
 (a) Transmission in the same local-exchange ensemble as
 Eq.~\eqref{eq:swapfloor}, for depth $r=2$, sixteen-cell blocks and
 $N=16384$. Curves are means of 256 paired independent samples; shading
 gives the 10th--90th realization percentiles, not confidence intervals.
 Each block makes at most one exchange attempt, with probability
 $p_{\rm sw}$ at a uniformly chosen internal boundary. All samples have
 $\cT_N(0)=1$. (b) Exact quadratic coefficients
 $\Gamma_2=\lim_{E\to0}\gamma/E^2$ and the graph lower bounds for two
 fixed recursion depths. The horizontal variable is probability per
 block; the two depths have different block lengths. Parameters are
 $h=2$, $g=1.2$, $t_{\rm off}=0.25$, and $t_q=e^q$ for
 $q\in\{-1,0,1\}$, with $K=1$ and matched contacts.}
 \label{fig:faults}
\end{figure*}

\emph{The localization rate and local exchanges.}---We measure energy $E$
from the resonance. At fixed energy and disorder law, the inverse localization length $\gamma(E)$ determines
typical transmission through a long chain according to
$\E\log\cT_N(E)=-2N\gamma(E)+o(N)$, where $N$ counts cells
and $\E$ averages over their random arrangement. Attenuation becomes
appreciable when $N\gamma$ is of order one. Hence
$\gamma\propto E^\nu$ suggests the scale $|E|\sim N^{-1/\nu}$:
a higher power permits a wider neighborhood at large $N$. Finite-device
bounds below make this scale precise without exchanging the two limits.

Each cell has one of finitely many types, labeled by $a$. Its transfer
matrix $M_a$ propagates the two-component wave amplitude across that cell.
We consider real analytic, determinant-one matrices satisfying
\begin{equation}
 M_a(0)=\diag(t_a/K,K/t_a),\qquad M_a'(0)=\frac{Q_a}{t_a}J,
 \label{eq:cellclass}
\end{equation}
where $t_a,Q_a>0$ and
$J=\left(\begin{smallmatrix}0&-1\\1&0\end{smallmatrix}\right)$.
The ratio $t_a/K$ sets the relative scaling of the two amplitudes at
resonance, while $Q_a/t_a$ sets their first-order mixing away from it.
In the quantum realization below, $t_a$ is the effective intracell hopping
and $K$ couples neighboring cells.
A block is an ordered sequence, or word, of $\ell$ cells. Its composition
$(n_a)$ specifies the number of cells of each type, and its transfer is
$P_w=M_{w_\ell}\cdots M_{w_1}$. Every type is present,
and at least two hopping values differ. Choosing the coupling scale
\begin{equation}
 K=\left(\prod_a t_a^{n_a}\right)^{1/\ell}
 \label{eq:balance}
\end{equation}
makes the transfer $P_w(0)$ of every allowed block equal to the identity.
At resonance the cell matrices are diagonal, so their product depends only
on the counts, not their order. Independent complete blocks drawn from any
probability distribution over these arrangements therefore give
$\cT_N(0)=1$ with matched contacts. Away from resonance, the off-diagonal
terms mix wave amplitudes, and their accumulated effect depends on order.

In each block draw a uniform internal boundary and, independently, attempt
its adjacent exchange with probability $p_{\rm sw}$. Equal neighbors give
no change. These choices are independent between blocks. We first
quantify the first-order change caused by exchanging types $a,b$,
then ask whether any arrangement can avoid all such changes. The local
quantity needed for this comparison is
\begin{equation}
 \chi_{ab}=\frac{|Q_a(t_b^2-K^2)-Q_b(t_a^2-K^2)|}{Kt_at_b}.
 \label{eq:chi}
\end{equation}
As shown below, this quantity is independent of the surrounding cells.
Connect the types by edges of weight $\chi_{ab}^2$. A block containing
all types traces a path connecting them, so its adjacent pairs cannot
have less total weight than the cheapest connected graph. Denote this
minimum spanning-tree weight by $W_{\rm tree}$. We prove
$W_{\rm tree}>0$ and the bound per physical cell
\begin{equation}
 \boxed{\gamma(E,p_{\rm sw})\ge
 \frac{p_{\rm sw}(1-p_{\rm sw})W_{\rm tree}}{2\ell(\ell-1)}
 E^2(1-C|E|).}
 \label{eq:swapfloor}
\end{equation}
For sufficiently small $|E|$, $C$ and the energy interval can be chosen
uniformly over all distributions of block arrangements with these cell
parameters and counts,
including probabilities depending on energy. Thus a fixed
$0<p_{\rm sw}<1$ forces quadratic localization even when ideal ordering
cancels it to higher order. The theorem keeps block length fixed; the
factor $\ell(\ell-1)$ records its resource dependence.

\emph{Why rearrangement cannot remove the scattering.}---For a balanced block,
the product $p_i=\prod_{j\le i}(t_{w_j}/K)$ records the zero-energy
rescaling through its first $i$ cells, with $p_0=p_\ell=1$.
The first derivative collects the linear mixing at each cell, weighted
by these rescalings:
\begin{equation}
 \begin{split}
 G_w=P_w'(0)&=\frac1K\begin{pmatrix}0&-A_w\\B_w&0\end{pmatrix},\\
 A_w&=\sum_i Q_{w_i}/p_i^2,\qquad
 B_w=\sum_iQ_{w_i}p_{i-1}^2.
 \end{split}
 \label{eq:derivative}
\end{equation}
If $w^{(j)}$ exchanges neighbors $a,b$, their derivative difference obeys
\begin{equation}
 -\det(G_{w^{(j)}}-G_w)=\chi_{ab}^2.                 \label{eq:swap}
\end{equation}
The cells preceding the exchanged pair rescale the individual matrix
entries, but their factors cancel in this determinant. The difference
matrix has trace zero and, when $\chi_{ab}>0$, real eigenvalues
$\pm\chi_{ab}$. It therefore contains a stretching component that
cannot be removed by a change of basis.

Some unequal pairs have $\chi_{ab}=0$ and cause no first-order change.
Nevertheless, every block contains every cell type, so its neighboring pairs
form a connected graph and their
total weight is at least $W_{\rm tree}$. Zero-weight edges join exactly
equal values of $(t_a^2-K^2)/Q_a$. The extreme hoppings lie on opposite
sides of the geometric mean $K$; zero-weight edges cannot connect the
whole set of types. Every spanning tree therefore has positive weight.

To turn this local change into a localization rate, separate the common
propagation from random scattering. Since $A_w,B_w>0$, the mean
first-order propagation can be put in a rotation basis. For the exchange
difference $D=G_{w^{(j)}}-G_w$, the squared wave-mixing coefficient in
this basis is $-\det D$ plus a nonnegative rotation contribution.
Thus the magnitude of the change in the wave-mixing amplitude is at
least $\chi_{ab}$.
Its independent yes-or-no choice contributes at least
$p_{\rm sw}(1-p_{\rm sw})\chi_{ab}^2$ to the variance of the
first-order reflection coefficient. The total variance equals twice the
quadratic localization coefficient per block~\cite{SchulzBaldes2007}.

The small-energy estimate must also resolve rare exchanges. In a basis
normalizing the exact mean transfer, let $\mathcal V(E)$ be the variance
of the block-to-block wave-mixing amplitude; it includes the energy dependence.
The Supplemental Material proves
$|\gamma_{\rm block}-\mathcal V(E)/2|\le C|E|\mathcal V(E)$.
Because the error retains the same small variance factor, it remains
controlled as the exchange probability tends to zero. Averaging over the
$\ell-1$ boundaries, applying the graph bound, and dividing by $\ell$
cells gives Eq.~\eqref{eq:swapfloor}.

\emph{A quantum chain with a high-transmission window.}---We next realize
the transfer class microscopically and show what ideal ordering gains.
Each cell has two core orbitals linked directly by $a$ and indirectly
through two auxiliary dimers. Each dimer has internal hopping $h$;
its two core--dimer couplings are $g+u_i$ and $g-u_i$, interchanged
between the two dimers. Neighboring cells couple through their core
orbitals with matrix element $-K$. Grouping the core orbitals and the
two auxiliary sets within each sublattice gives the one-particle Hamiltonian
\begin{equation}
 H_N=\begin{pmatrix}0&T_N\\T_N^\dagger&0\end{pmatrix},\qquad
 T_N=\begin{pmatrix}
 aI-KS_N&G_+&G_-\\G_-&hI&0\\G_+&0&hI
 \end{pmatrix},                                     \label{eq:model}
\end{equation}
Here $S_N$ connects neighboring cells with open boundaries,
$G_\pm=\diag(g\pm u_i)$, and $K,h>0$. Eliminating the auxiliary
amplitudes leaves an exact two-core recurrence. Its energy-dependent
hopping $t_i(E)$ includes the two indirect paths, while $z_i(E)$ is
the energy shifted by their local self-energy:
\begin{align}
 z_i(E)&=E\left[1+\frac{2(g^2+u_i^2)}{h^2-E^2}\right],\nonumber\\
 t_i(E)&=a-\frac{2h(g^2-u_i^2)}{h^2-E^2},\quad |E|<h, \label{eq:energy}\\
 M_i(E)&=\begin{pmatrix}t/K-z^2/(Kt)&-z/t\\z/t&K/t\end{pmatrix},
                                                     \label{eq:transfer}
\end{align}
where $t=t_i(E)$ and $z=z_i(E)$. These formulas retain the full energy
dependence of the auxiliary orbitals. At zero energy
$t_i=t_{\rm off}+2u_i^2/h>0$, with $t_{\rm off}=a-2g^2/h$, and
$Q_i=z_i'(0)=1+2(g^2+u_i^2)/h^2$, which identifies the mixing
coefficient in Eq.~\eqref{eq:cellclass}. The affine relation $Q(t)=\alpha+t/h$,
$\alpha=1+2g^2/h^2-t_{\rm off}/h$, makes every unequal exchange active
throughout the physically allowed parameter family.

For three positive levels $t_-<t_0<t_+$, take the reverse four-cell
seeds $w_A=(-,0,0,+)$ and $w_B=(+,0,0,-)$; the symbols label hopping
levels. A common calibration change $t\mapsto F(t)>0$, applied to all cells,
preserves their composition balance after retuning intercell and
contact hoppings to
\begin{equation}
 K_F=[F(t_-)F(t_0)^2F(t_+)]^{1/4}.                   \label{eq:retune}
\end{equation}
The calibrated levels are realized by
$u_j^2=(h/2)[F(t_j)-t_{\rm off}]\ge0$.
This composition protection is related to correlated-block mechanisms
\cite{Ammari2025}. A common offset can instead change $a$ with the coupling imbalances
$u_i$ fixed. In both cases the complete-device central law is
\cite{Kasturirangan2022}
\begin{equation}
 \cT_N(0)=\operatorname{sech}^2\sum_i\log(t_i/K).      \label{eq:zero}
\end{equation}

Let $A_0=P_{w_A}$ and $B_0=P_{w_B}$. The established multipolar
recursion forms the two opposite orders of the same pair of blocks:
$A_{r+1}=A_rB_r$, $B_{r+1}=B_rA_r$. At fixed depth $r$, choose
independently and fairly between the two length-$\ell_r=4\,2^r$ blocks.
Doubling the block suppresses their difference because
$A_{r+1}-B_{r+1}=[A_r,B_r]$ is a commutator. Since both transfers approach the identity,
an $O(E^{r+1})$ difference becomes $O(E^{r+2})$ after this step.
Starting from the seeds' linear difference, the random transfer amplitude
thus scales as $E^{r+1}$. Its variance sets the localization rate,
giving the doubled power. The nonzero coefficient and the vanishing
mean rotation angle are treated in the Supplemental Material using
Refs.~\cite{SchulzBaldes2007,Mo2025}. We obtain
\begin{equation}
 \gamma_r(E)=a_r E^{2r+2}+o(E^{2r+2}),\qquad a_r>0.   \label{eq:lyapunov}
\end{equation}
Every feasible strictly monotone common remapping preserves this power.
Strict positivity follows from the ordered prefixes of $w_A$ in
Eq.~\eqref{eq:derivative}; the map changes the coefficient, not its order.

\begin{figure*}[t]
 \includegraphics[width=0.95\textwidth]{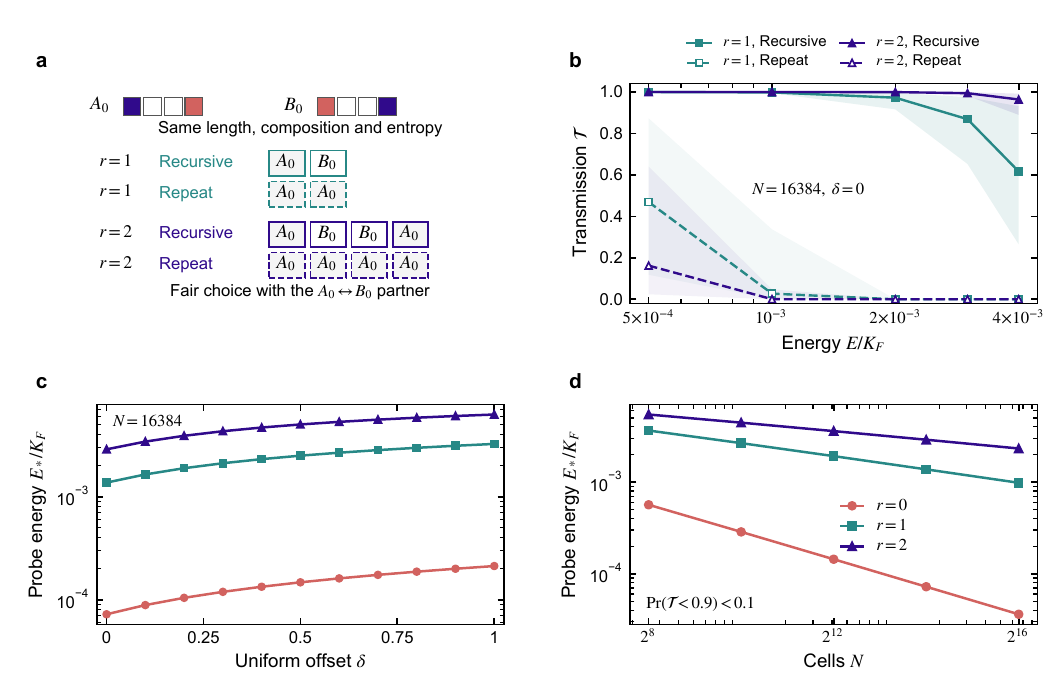}
 \caption{Ordering controls transmission at equal resources.
 (a) Balanced seeds, recursive words and repeated-seed controls; purple,
 white and red denote $t_-,t_0,t_+$. Each depth has equal block length,
 amplitude distribution and entropy across the two ordering schemes.
 (b) Transmission at $N=16384$, zero offset: medians and 10--90\%
 realization bands from 256 sequences paired across ordering schemes and
 energies. (c,d) Positive energies guaranteed to satisfy
 $\Pr(\cT<0.9)<0.1$ versus common offset and length. Each marker has an
 interval-arithmetic enclosure; connecting lines guide the eye.
 Parameters are those of Fig.~\ref{fig:faults}. The offset changes $a$
 at fixed coupling imbalances $u_i$, with intercell and contact hoppings retuned together.}
 \label{fig:windows}
\end{figure*}

To isolate ordering, compare with fair repeated seeds
$A_0^{2^r},B_0^{2^r}$. They have identical block length, composition,
amplitude distribution and entropy $\log2/\ell_r$ per cell, yet
$\gamma_{{\rm repeat},r}(E)=b_rE^2+o(E^2)$, with explicit $b_r>0$.
For sixteen-cell blocks, $N=16384$ and $E=0.002$, median transmissions
are $0.99948$ and $5.26\times10^{-15}$ over 256 paired sequences
[Fig.~\ref{fig:windows}(b)]. Exact second-moment bounds further give
pointwise high-transmission guarantees at $|E|\sim N^{-1/(2r+2)}$;
Figs.~\ref{fig:windows}(c,d) show energies satisfying these bounds.
The comparison demonstrates the transmission gained by arranging the same
cells in a different order.

\emph{Local errors in a finite device.}---Figure~\ref{fig:faults}(a)
applies the theorem's adjacent exchanges directly to the depth-two
construction. The central transmission remains one at all error rates,
whereas the nearby curves separate. The restored quadratic term does not
erase the useful transmission of a finite device immediately.
For the same exchange distribution, the mean excess resistance
$\rho=\cT^{-1}-1$ follows exactly from a $4\times4$ second-moment
transfer map. At $N=16384$ and $E=0.002$, raising $p_{\rm sw}$
from zero to $0.01$ increases $\E\rho$ from about $0.000792$ to $0.009900$.
Jensen's inequality then gives $\E\cT\ge1/(1+\E\rho)\simeq0.9902$.
Thus appreciable transmission can persist at finite length even after
the asymptotic localization power has changed.

\emph{Tolerance to block replacement.}---A different assembly error gives
an explicit relation between error rate and useful device length.
Replace an ideal recursive word by a repeated-seed word of the same
length and composition with probability $p_{\rm rep}$, choosing the two
orientations fairly in either class. All samples still have $\cT_N(0)=1$,
and uniformly in $p_{\rm rep}$,
\begin{equation}
 \begin{split}
 \gamma_r(E,p_{\rm rep})={}&(1-p_{\rm rep})a_rE^{2r+2}
                         +p_{\rm rep}b_rE^2\\
 &+o(|E|^{2r+2}+p_{\rm rep}E^2).
 \end{split}                                       \label{eq:faultgamma}
\end{equation}
At an ideal-window energy $E=bN^{-1/(2r+2)}$ with fixed $b>0$, the
ideal contribution $NE^{2r+2}$ stays finite, whereas the replacement
contribution accumulates as $p_{\rm rep}NE^2=p_{\rm rep}b^2N^{r/(r+1)}$.
This identifies the expected error scale, but a typical-transmission
exponent alone does not bound a mean resistance. Independent upper and
lower bounds on the exact finite-product second moment prove that
bounded mean excess resistance at this energy is equivalent to
\begin{equation}
 \begin{gathered}
 p_{\rm rep}N^{r/(r+1)}=O(1)\\
 \text{(four-word replacement ensemble)}.
 \end{gathered}                                     \label{eq:budget}
\end{equation}
Sufficiently small energy and error-rate prefactors additionally give
high transmission probability. The Supplemental Material derives the standard one-channel
resistance diffusion and its microscopic scale
\cite{Landauer1970,Beenakker1997,Bachmann2012}. This finite-length equivalence concerns block replacement; neighboring
exchanges obey the general localization bound in Eq.~\eqref{eq:swapfloor}.

The wider ideal window also retains a mean-tuning cost. A uniform error
$\eta$ in the retuned $\log K$, with contacts matched to the implemented
value, gives
\begin{equation}
 \cT_N(0)=\operatorname{sech}^2(N\eta),\qquad
 |\eta|\le\frac{\operatorname{atanh}\sqrt\varepsilon}{N}
 \label{eq:precision}
\end{equation}
for $\cT_N(0)\ge1-\varepsilon$. At $N=16384$ and $\varepsilon=0.1$
the tolerance is about 20 parts per million in $\log K$.
Mean calibration and ordering reliability control distinct failure modes.

\emph{A measurable distinction.}---Frequency-resolved microwave
transmission and spatially resolved optical propagation already provide
experimental access to transport in correlated disorder~\cite{Kuhl2000,Naether2013}.
A reciprocal network realizing the coupled-mode Hamiltonian in
Eq.~\eqref{eq:model} could test our prediction by measuring the two-port
flux-normalized transmission $\cT=|S_{21}|^2$ before and after neighboring exchanges.
In the lossless, matched-contact limit the central peak is unchanged,
while the surrounding spectra differ. Measurements at several lengths
would estimate $\gamma$ from half the long-length slope of $-\E\log\cT_N$ and distinguish
the restored quadratic term from finite-sample interference.

The number of each cell type protects the resonance; their order controls
how far this protection extends in energy. The local scattering invariant
shows why no fixed block arrangement in the stated class can remove every
exchange-induced scattering channel. This separates robustness of the peak
from robustness of transport around it, and supplies a test through the
transmission spectrum rather than through microscopic defect identification.

\emph{Acknowledgments}---This work is supported by Natural Science Foundation of Hebei Province (Grant No. A2026203033), Beijing National Laboratory for Condensed Matter Physics (Grant No. 2024BNLCMPKF020).

\bibliography{paper2}

\end{document}


\maketitle
Composition balance fixes the block transfer exactly at zero energy, including
after common calibration and retuning. To determine nearby transmission,
we track the order-dependent terms in its energy expansion. Recursive
ordering pushes the random part of the transfer to higher orders; a local
swap can restore linear fluctuations.
The Lyapunov estimates
convert these fluctuations into bulk localization rates. Separate
finite-product estimates then control device transmission, and the
four-word replacement ensemble supplies a microscopic length scale for
the resistance diffusion. The latter describes the length process at each
specified small energy.
\section{Finite-energy transfer of the decorated chain}
\label{sm:energy}

The microscopic Hamiltonian in the main text has an exact
energy-dependent core transfer matrix. For $|E|<h$, elimination of the local
mediator dimers gives
\begin{align}
 z_i(E)&=E\left[1+\frac{2(g^2+u_i^2)}{h^2-E^2}\right],&
 t_i(E)&=a-\frac{2h(g^2-u_i^2)}{h^2-E^2},
 \label{sm:energy:elimination}\\
 z_i(E)L_i&=t_i(E)R_i-KR_{i-1},&
 z_i(E)R_i&=t_i(E)L_i-KL_{i+1}.
\end{align}
Thus $(L_{i+1},R_i)^T=M_i(E)(L_i,R_{i-1})^T$, where
\begin{equation}
 M_i(E)=\begin{pmatrix}
 t_i(E)/K-z_i(E)^2/[Kt_i(E)]&-z_i(E)/t_i(E)\\
 z_i(E)/t_i(E)&K/t_i(E)
 \end{pmatrix},\qquad \det M_i(E)=1.
 \label{sm:energy:cell}
\end{equation}
The finitely many $t_i(0)=t_i$ are positive, so these matrices are real
analytic in a common neighborhood of zero. No large-$h$ expansion is
required. Set
\begin{equation}
 Q_i=1+\frac{2(g^2+u_i^2)}{h^2},\quad
 r_i=\frac{t_i}{K},\quad c_i=\frac{t_i-a}{h^2},\quad
 J=\begin{pmatrix}0&-1\\1&0\end{pmatrix}.
\end{equation}
Here $Q_i=z_i'(0)>0$ controls the linear mixing of the core amplitudes. Expansion of
Eq.~\eqref{sm:energy:cell} gives
\begin{equation}
 M_i(E)=\begin{pmatrix}r_i&0\\0&r_i^{-1}\end{pmatrix}
 +E\frac{Q_i}{t_i}J
 +E^2\begin{pmatrix}c_i/K-Q_i^2/(Kt_i)&0\\0&-Kc_i/t_i^2\end{pmatrix}
 +O(E^3).
 \label{sm:energy:cell-expansion}
\end{equation}

For a word $w=(1,\ldots,L)$, let
$P_w(E)=M_L(E)\cdots M_1(E)$. Suppose that $\prod_i r_i=1$, and
write $p_i=\prod_{j\le i}r_j$, $p_0=p_L=1$. Differentiating the product
and using its diagonal zero-energy factors yields
\begin{equation}
 P_w(0)=I,\qquad P_w'(0)=\frac1K\begin{pmatrix}0&-A_w\\B_w&0\end{pmatrix},
 \qquad A_w=\sum_i\frac{Q_i}{p_i^2},\quad
 B_w=\sum_i Q_i p_{i-1}^2.
 \label{sm:energy:word-derivative}
\end{equation}
Reversing a word interchanges $A_w$ and $B_w$. Composition balance
therefore fixes the zero-energy transfer but leaves an order-dependent
first derivative.

\section{Recursive ordering and the localization exponent}
\label{sm:energy:hierarchy}

Choose three positive hopping levels and the reverse seed words
$w_A=(-,0,0,+)$ and $w_B=(+,0,0,-)$. Their common critical value is
$K=(t_-t_0^2t_+)^{1/4}$. Write their transfers as $A_0(E),B_0(E)$ and
their derivatives as
\begin{equation}
 A_0'(0)=\begin{pmatrix}0&-a_*\\b_*&0\end{pmatrix},\qquad
 B_0'(0)=\begin{pmatrix}0&-b_*\\a_*&0\end{pmatrix},\qquad
 \omega_0=\frac{a_*+b_*}{2},\quad c_0=|a_*-b_*|.
 \label{sm:energy:seeds}
\end{equation}
Both $a_*,b_*$ are positive. Define the matrix recursion
\begin{equation}
 A_{r+1}=A_rB_r,\qquad B_{r+1}=B_rA_r.
 \label{sm:energy:recursion}
\end{equation}
Physical spatial words concatenate in the reverse order because their
transfer matrices multiply from right to left; this only exchanges the
two word labels. The depth-$r$ word length is $L_r=4\,2^r$. At each
fixed depth, the chain consists of independent fair choices of these
two words.

The recursion and its cancellation of low-order expansion coefficients
have direct precedents in random multipolar driving
\cite{Zhao2021,Mori2021}. Higher-order localization anomalies also occur
in correlated random chains \cite{Giri1993,Jitomirskaya2003}.
The derivation below treats the real noncompact transfer matrices at a
common identity resonance, where their rotation angle tends to zero.
The normalized-mean mechanism and high-order growth have a direct
noncompact precedent in Ref.~\cite{Mo2025}; the formulas here retain the
microscopic coefficients and matched spatial contacts
\cite{SchulzBaldes2007}.

\paragraph{Cancellation of transfer derivatives.}
If $c_0>0$, then for every fixed $r\ge0$,
\begin{align}
 \frac{A_r'(0)+B_r'(0)}2&=\omega_rJ,\qquad \omega_r=2^r\omega_0,\\
 A_r(E)-B_r(E)&=E^{r+1}C_r+O(E^{r+2}),
 \label{sm:energy:jet}\\
 C_{r+1}&=[C_r,\omega_rJ],\qquad
 c_r=c_0\omega_0^r2^{r(r+1)/2}.
 \label{sm:energy:c-recursion}
\end{align}
Here $C_r$ is symmetric and traceless, with eigenvalues $\pm c_r$.
Indeed, if $\overline A_r=(A_r+B_r)/2$ and $\Delta_r=A_r-B_r$, then
$A_rB_r-B_rA_r=[\Delta_r,\overline A_r]$. Its commutator with the
constant term $I$ vanishes, while its next coefficient is
$[C_r,\omega_rJ]$. Commutation with $\omega_rJ$ maps a symmetric
traceless matrix with eigenvalues $\pm c_r$ to one with eigenvalues
$\pm2\omega_r c_r$. This proves the induction, including the coefficient.

\subsection{A Lyapunov lemma at a vanishing rotation angle}
\label{sm:energy:lyapunov}

Let $M_\pm(E)$ be real analytic $SL(2,\mathbb R)$ matrices satisfying
$M_\pm(0)=I$. Suppose their mean derivative is $H$, with
$H^2=-\omega^2 I$, $\omega>0$, and
\begin{equation}
 M_+(E)-M_-(E)=E^pC_p+O(E^{p+1}),\qquad p\ge1.
 \label{sm:energy:binary-assumption}
\end{equation}
Choose a real determinant-one basis change $Q_0$ such that
$Q_0HQ_0^{-1}=\omega J$, allowing the opposite orientation of $J$ if
needed. Let $S$ be the symmetric traceless part of
$Z=Q_0C_pQ_0^{-1}/2$, and let its eigenvalues be $\pm s$.
For independent fair choices of the two matrices, the top Lyapunov
exponent per matrix satisfies
\begin{equation}
 \gamma(E)=\frac{s^2}{2}E^{2p}+o(E^{2p}).
 \label{sm:energy:binary-lyapunov}
\end{equation}
The formula permits $s=0$, in which case it specifies a vanishing
coefficient rather than a positive leading term.

\paragraph{Mean normalization.}
The arithmetic mean need not preserve determinant one, even though each
transfer does. Normalizing its determinant keeps its scalar stretch
explicit while isolating the common elliptic motion, so rotation and
fluctuating stretching can be estimated separately.
Put $C=(M_++M_-)/2$, $\Delta=M_+-M_-$, and $D=C^{-1}\Delta/2$.
For sufficiently small $E$, $\det C>0$. The two unit-determinant
conditions give exactly
\begin{equation}
 \operatorname{tr}D=0,\qquad \det C\,(1+\det D)=1.
 \label{sm:energy:mean-det}
\end{equation}
The normalized mean $\widehat C=C/\sqrt{\det C}$ lies in $SL(2,\mathbb R)$
and has derivative $H$ at zero. Consequently
$\operatorname{tr}\widehat C=2-\omega^2E^2+O(E^3)$, so it is elliptic
for every sufficiently small nonzero real $E$. There is a bounded real
basis change $Q(E)\to Q_0$ with
$Q\widehat C Q^{-1}=R_{\theta(E)}$,
$\theta(E)=\omega E+O(E^2)$.
For example,
$(\widehat C-\tfrac12\operatorname{tr}\widehat C\,I)/\sin\theta$
extends to $H/\omega$, squares to $-I$, and determines an invariant
positive quadratic form and hence such a locally analytic conjugation.
In this basis,
\begin{equation}
 \widetilde M_\pm=
 \frac{R_\theta(I\pm\widetilde D)}{\sqrt{1+\det D}},\qquad
 \widetilde D=QDQ^{-1}=E^pZ+O(E^{p+1}).
 \label{sm:energy:rotating-frame}
\end{equation}

\paragraph{Stationary projective measure.}
For a twice continuously differentiable function $f$ of projective
angle, the averaged transition satisfies uniformly in angle
\begin{equation}
 P_Ef-f=\theta(E)f'+O(E^2+\|\widetilde D\|^2).
 \label{sm:energy:stationarity}
\end{equation}
The linear contributions of $+\widetilde D$ and $-\widetilde D$ cancel.
Let $\nu_E$ be any stationary projective measure. Integrating
Eq.~\eqref{sm:energy:stationarity}, dividing by $\theta(E)$, and passing
to a weakly convergent subsequence gives $\int f'\,d\nu=0$ because
$(E^2+E^{2p})/|\theta(E)|\to0$. Every limiting measure is therefore
uniform on the projective circle. This explicitly handles the
zero-angle limit instead of assuming an angle bounded away from zero.

\paragraph{Radial growth.}
For a unit vector $v$, the fair-sign mean logarithmic stretch is
\begin{align}
 h_E(v)={}&-\frac12\log(1+\det D)\nonumber\\
 &+\frac14\log\left[
 (1+2v^T\widetilde Dv+\|\widetilde Dv\|^2)
 (1-2v^T\widetilde Dv+\|\widetilde Dv\|^2)\right].
\end{align}
Its uniform leading term is
\begin{equation}
 \frac{h_E(v)}{E^{2p}}\longrightarrow
 -\frac12\det Z+\frac12\|Zv\|^2-(v^TZv)^2.
 \label{sm:energy:radial}
\end{equation}
Apply the stationary formula for the top Lyapunov exponent to a
stationary measure realizing that exponent. Uniqueness of the stationary
measure is unnecessary, since every weak limit was fixed above.
Writing $Z=\zeta J+S$, uniform angular averages give
\begin{equation}
 \det Z=\zeta^2-s^2,\quad
 \langle\|Zv\|^2\rangle=\zeta^2+s^2,\quad
 \langle(v^TZv)^2\rangle=s^2/2.
\end{equation}
Their substitution in Eq.~\eqref{sm:energy:radial} proves
Eq.~\eqref{sm:energy:binary-lyapunov}. In particular, no inverse power
of $E$ is lost through the vanishing angular drift.

\subsection{Localization coefficients and the fixed-composition reference}

For the reverse seeds, $Q_0=I$, $p=r+1$, and $Z=C_r/2$. Thus
\begin{align}
 \gamma_r^{\rm block}(E)&=\frac{c_r^2}{8}E^{2r+2}+o(E^{2r+2}),
 \label{sm:energy:block-exponent}\\
 \gamma_r^{\rm cell}(E)&=
 \frac{c_0^2\omega_0^{2r}2^{r^2}}{32}E^{2r+2}+o(E^{2r+2}).
 \label{sm:energy:cell-exponent}
\end{align}
The latter divides by $L_r=4\,2^r$. The first three depths therefore
have quadratic, quartic, and sixth-order localization, with all
finite-mediator factors retained.

This suppression is not a consequence of composition balance alone.
If all 12 permutations of $(-,0,0,+)$ are chosen independently and
uniformly, reversal symmetry makes their mean first derivative
$\omega J$. Write each first derivative as $G_w=s_wJ+t_wX$, where
$X=\left(\begin{smallmatrix}0&1\\1&0\end{smallmatrix}\right)$ and
$t_w=(B_w-A_w)/(2K)$. For
$P_w=I+EG_w+E^2H_w+O(E^3)$, the determinant condition gives
$\operatorname{tr}H_w=-\det G_w=t_w^2-s_w^2$.
The first-order averaged radial stretch vanishes, the stationary angle
again converges to the uniform measure, and the second-order radial
average is $t_w^2/2$. Consequently
\begin{equation}
 \gamma_{12}^{\rm block}(E)=\frac{E^2}{8K^2}
 \mathbb E_w(A_w-B_w)^2+o(E^2).
 \label{sm:energy:unrestricted}
\end{equation}
For distinct ordered levels the coefficient is positive by the seed
inequality proved below.

\section{Finite-length transmission certificate}
\label{sm:energy:certificate}

The Lyapunov exponent describes infinite-length logarithmic growth at
fixed energy; it does not itself bound the loss probability of a finite
device whose energy also decreases with length. We therefore control the
finite product's second moment directly. Set
$A=(M_++M_-)/2$, $D_0=(M_+-M_-)/2$, and use the mean-rotation basis
$Q(E)$ above. A prime denotes conjugation by $Q$. Independence of the
fair signs gives the exact identities
\begin{equation}
 \mathbb E(M_\sigma'^TM_\sigma')
 =\det A\,I+D_0'^TD_0',\qquad
 \det A+\det D_0=1.
 \label{sm:energy:second-moment}
\end{equation}
Define the nonnegative, directly computable quantity
\begin{equation}
 q(E)=\|D_0'\|_2^2-\det D_0\ge0.
 \label{sm:energy:q}
\end{equation}
Then $\mathbb E(M_\sigma'^TM_\sigma')\preceq[1+q(E)]I$.
For the product $P_n'$ of $n$ independent depth-$r$ superblocks,
conditioning on one factor at a time yields
\begin{equation}
 \mathbb E\|P_n'\|_F^2\le2[1+q(E)]^n,
 \qquad q(E)=\frac{c_r^2}{2}E^{2r+2}+O(E^{2r+3}).
 \label{sm:energy:finite-product}
\end{equation}
In particular, $q(E)\le C_r^*|E|^{2r+2}$ in a sufficiently small
energy interval. The limit of its leading coefficient is explicit;
the exact form \eqref{sm:energy:q} avoids estimating a Taylor remainder.

\subsection{Flux normalization and contacts}

Attach identical uniform leads of hopping $K$, with contact hopping $K$,
to the two outer core orbitals. This specifies matching at finite energy,
not just the product of zero-energy contact broadenings. In the same
core coordinates the lead cell transfer is
\begin{equation}
 M_\ell(E)=\begin{pmatrix}1-x^2&-x\\x&1\end{pmatrix},\qquad x=E/K.
\end{equation}
For $|x|<2$, its determinant-one invariant metric is
\begin{equation}
 G_\ell(E)=\frac1{\sqrt{1-x^2/4}}
 \begin{pmatrix}1&x/2\\x/2&1\end{pmatrix},\qquad
 M_\ell^TG_\ell M_\ell=G_\ell.
 \label{sm:energy:lead-metric}
\end{equation}
Choose $L(E)=G_\ell(E)^{1/2}$, so $L(0)=I$ and
$LM_\ell L^{-1}$ is a rotation. For the device transfer $P_n$, the
transmission is exactly
\begin{equation}
 \mathcal T_n(E)=\frac4{\|L(E)P_n(E)L(E)^{-1}\|_F^2+2}.
 \label{sm:energy:transmission}
\end{equation}
To see this, change the real rotation coordinates to the two complex
traveling-wave amplitudes. The resulting transfer is in $SU(1,1)$,
with transmission $1/|\alpha|^2$ and Frobenius norm squared
$4|\alpha|^2-2$. Endpoint phase rotations do not affect that norm.
The elimination \eqref{sm:energy:elimination} preserves the intercell
core current, so no additional mediator metric multiplies this relation.

Put $k(E)=\operatorname{cond}_2[L(E)Q(E)^{-1}]$. Reverse seeds have
$Q(0)=L(0)=I$, hence $k(E)=1+O(|E|)$. Equations
\eqref{sm:energy:finite-product} and \eqref{sm:energy:transmission} imply
\begin{align}
 \mathbb E[\mathcal T_n(E)^{-1}-1]
 &\le\frac{k(E)^2[1+q(E)]^n-1}{2},
 \label{sm:energy:inverse-bound}\\
 \mathbb E\mathcal T_n(E)
 &\ge\frac2{1+k(E)^2[1+q(E)]^n},
 \label{sm:energy:mean-bound}\\
 \Pr\{\mathcal T_n(E)<1-\varepsilon\}
 &\le\frac{1-\varepsilon}{2\varepsilon}
 \left[k(E)^2[1+q(E)]^n-1\right],\quad0<\varepsilon<1.
 \label{sm:energy:probability-bound}
\end{align}
The second inequality is Jensen's inequality for the inverse; the third
is Markov's inequality applied to $\mathcal T^{-1}-1$.

For a prescribed loss threshold and failure probability, choose a small
constant $b>0$ and take $|E|\le b\,n^{-1/(2r+2)}$. Since
$[1+q(E)]^n\le\exp(C_r^*n|E|^{2r+2})$ and $k(E)\to1$, the failure
bound is below the prescribed probability for all sufficiently large
$n$. This is a finite-device energy window, with physical length
$N=L_rn$, proved independently of the Lyapunov limit. It is a bound at
each specified energy, not a claim of simultaneous high transmission
throughout a continuum of energies in every realization. A finite set
of energies can be treated by a union bound.

\section{Common calibration, tuning accuracy, and resource costs}
\label{sm:energy:calibration}

For any common positive map $F$ on the three hopping levels, retune
\begin{equation}
 K_F=[F(t_-)F(t_0)^2F(t_+)]^{1/4}.
 \label{sm:energy:retuning}
\end{equation}
The two seeds and every superblock have identical composition, so their
zero-energy transfers remain exactly $I$. The derivative cancellation
\eqref{sm:energy:jet} therefore survives a common calibration without
knowing $F$ pointwise. Its finite-energy coefficients change.

For strictly increasing $F$, the transformed ratios obey
$r_-<r_0<r_+$ and $r_-r_0^2r_+=1$. The interior prefix products of
the ordered seed satisfy
$r_-<1$, $r_-r_0<1$, and $r_-r_0^2=1/r_+<1$. It follows that
\begin{equation}
 A_w-B_w=\sum_iQ_i(p_i^{-2}-p_{i-1}^2)>0.
 \label{sm:energy:strict-coefficient}
\end{equation}
The decreasing case follows by reversal. Hence $c_0>0$ for every
strictly monotone feasible calibration, including all feasible additive
offsets, and the exponent in Eq.~\eqref{sm:energy:cell-exponent} is
exactly $2r+2$. For arbitrary nonmonotone positive maps the cancellation
still holds, but its leading coefficient can vanish. A constant map
removes all disorder.

The six-orbital realization requires real sources. For example, choose
$t_{\rm off}<\min_aF(t_a)$ and set
$u_i^2=(h/2)[F(t_i)-t_{\rm off}]$. A common additive offset can instead
be implemented by shifting $a$ at fixed sources. At fixed depth, a
compact feasible calibration family with positive hopping lower bounds,
bounded microscopic parameters, and $h$ bounded away from zero admits
uniform constants and a common small-energy interval in
Eqs.~\eqref{sm:energy:finite-product}--\eqref{sm:energy:probability-bound}.
For a compact family of distinct ordered transformed levels the leading
coefficient is also uniformly positive.

Near-unit transmission in Eq.~\eqref{sm:energy:probability-bound}
assumes lead and contact matching at the retuned value $K_F$. With
fixed unmatched contacts, a bounded scattering-basis change still gives
a positive finite-length transmission bound, but its zero-energy
contact loss need not vanish. Bulk retuning and contact matching are
separate resources.

If the implemented intercell hopping is $K_{\rm used}=K_Fe^\eta$, with
contacts matched to $K_{\rm used}$, every complete balanced sample of
$N$ cells has residual mass $-N\eta$. The exact zero-energy law gives
\begin{equation}
 \mathcal T_N(0)=\operatorname{sech}^2(N\eta),\qquad
 \mathcal T_N(0)\ge1-\varepsilon
 \ \Longleftrightarrow\
 |\eta|\le\frac{\operatorname{artanh}\sqrt\varepsilon}{N}.
 \label{sm:energy:detuning}
\end{equation}
Thus the energy window widens to $N^{-1/(2r+2)}$ at fixed depth, whereas
the required accuracy of the scalar critical retuning remains $N^{-1}$.
The construction tolerates a common calibration after retuning; it does
not determine that critical setting without measurement.

Finally, the fair depth-$r$ block sequence has entropy
$\log2/(4\,2^r)$ per cell. Increasing depth doubles its block length
and increases the coefficient in Eq.~\eqref{sm:energy:cell-exponent}.
All asymptotic statements hold with $r$ fixed. They neither justify
taking $r\to\infty$ with $N$ nor establish a finite interval of extended
states. Independent cell errors and arbitrary assembly errors are outside
the common-calibration assumption. The focused supplement separately
treats independent local transpositions and the specified four-word
replacement ensemble.

\section{A lower bound for local exchanges in composition-balanced blocks}
\label{sm:swap}

Consider a finite alphabet of positive hoppings $t_a$ and positive
cell metrics $Q_a$. The exact cell transfers are real analytic near
$E=0$, belong to $\mathrm{SL}(2,\mathbb R)$, and obey
\begin{equation}
 M_a(0)=\operatorname{diag}(t_a/K,K/t_a),\qquad
 M_a'(0)=\frac{Q_a}{t_a}J,\qquad
 J=\begin{pmatrix}0&-1\\1&0\end{pmatrix}.
 \label{sm:swap:cell}
\end{equation}
Fix the block length $\ell$. For a word $w$, put
$x_i=t_{w_i}/K$, $p_i=\prod_{j\le i}x_j$, and $p_0=1$.
Use $P_w=M_{w_\ell}\cdots M_{w_1}$. A balanced word has $p_\ell=1$,
hence $P_w(0)=I$. Direct differentiation yields
\begin{equation}
 G_w=P_w'(0)=\frac1K\begin{pmatrix}0&-A_w\\B_w&0\end{pmatrix},
 \quad A_w=\sum_{i=1}^{\ell}\frac{Q_{w_i}}{p_i^2},\quad
 B_w=\sum_{i=1}^{\ell}Q_{w_i}p_{i-1}^2.
 \label{sm:swap:derivative}
\end{equation}
In particular $A_w,B_w>0$. Every probability mixture of balanced
words therefore has an elliptic mean first derivative. All its
normalizers and rotation speeds lie in a compact, nondegenerate set
when the alphabet and block length are fixed.

If the words share a composition $(n_a)$, the single retuning
$K=(\prod_a t_a^{n_a})^{1/\ell}$ balances all permutations.
This remains true after a common positive map of the alphabet, with
$K$ retuned to its new geometric mean. The conclusions below apply
at each feasible calibration satisfying Eq.~\eqref{sm:swap:cell}.

\subsection{The local invariant}

A swap can change the derivative while preserving the zero-energy product.
We first isolate a measure of this change that survives the basis
transformation needed for localization.

Swap neighboring hoppings $a,b$ after a prefix product $P$, and write
$x=a/K$, $y=b/K$. With swapped minus original as the sign convention,
Eq.~\eqref{sm:swap:derivative} gives
\begin{align}
 D_{w,j}&=G_{w^{(j)}}-G_w
 =\begin{pmatrix}0&F_{ab}/(KP^2x^2y^2)\\P^2F_{ab}/K&0\end{pmatrix},
 \nonumber\\
 F_{ab}&=Q_a(y^2-1)+Q_b(1-x^2),\qquad
 -\det D_{w,j}=\chi_{ab}^2,
 \label{sm:swap:invariant}\\
 \chi_{ab}&=\frac{|Q_a(b^2-K^2)-Q_b(a^2-K^2)|}{Kab}.
 \nonumber
\end{align}
The determinant does not depend on the prefix or swap position.
When $\chi_{ab}>0$, the derivative difference is hyperbolic in every
real basis. Some unequal pairs can have $\chi_{ab}=0$ in the general
positive-metric class.

For the finite-mediator model the metric is
\begin{equation}
 Q(t)=\alpha+t/h,\qquad
 \alpha=1+2g^2/h^2-t_{\rm off}/h,
 \quad t=t_{\rm off}+2u^2/h.
\end{equation}
Thus
\begin{equation}
 \chi_{ab}=\frac{|b-a|}{Kab}
 \left[\alpha(a+b)+\frac{ab+K^2}{h}\right].
 \label{sm:swap:physical}
\end{equation}
Let $t_* =\min_a t_a$. Feasibility gives $Q(t_*)\ge1$, and balancing
gives $K\ge t_*$. The bracket in Eq.~\eqref{sm:swap:physical} equals
\begin{equation}
 Q(t_*)(a+b)+
 \frac{(a-t_*)(b-t_*)+K^2-t_*^2}{h}>0.
\end{equation}
Every unequal swap is consequently active in this model, including
parameter regimes with $\alpha<0$.

\section{An elliptic reflection estimate uniform in the probabilities}
\label{sm:swap:reflection}

The leading fixed-law expansion is the first-degree elliptic anomaly
of Ref.~\cite{SchulzBaldes2007}. We give a relative-error argument
because an absolute $O(E^3)$ remainder cannot control arbitrarily
rare faults.
Let $P_\sigma=I+O(E)$ range over a fixed finite analytic family with
$\det P_\sigma=1$ and uniformly nondegenerate elliptic mean derivative.
Its probabilities may depend on $E$. Put $C=\mathbb E P_\sigma$ and
choose a bounded real normalizer $R(E)$ so that
\begin{equation}
 RCR^{-1}=a_0\mathcal R_\theta,\quad
 a_0=\sqrt{\det C}>0,\quad
 \theta=\Omega E+O(E^2),\quad 0<\Omega_-\le\Omega\le\Omega_+.
 \label{sm:swap:meanbasis}
\end{equation}
Such a normalizer follows by normalizing the elliptic matrix
$C/\sqrt{\det C}$ and differs by $O(|E|)$ from the normalizer of
$\mathbb E G_\sigma$. For energy-dependent probabilities this expectation
is the weighted first jet at each energy, not a total derivative of
$C(E)$ along the probability law; the latter need not exist.
For $L_\sigma=RP_\sigma R^{-1}=(l_{ij})$ define the complex action
$v\mapsto a_\sigma v+b_\sigma\bar v$ by
\begin{equation}
 a_\sigma=\frac{l_{11}+l_{22}+i(l_{21}-l_{12})}{2},\qquad
 b_\sigma=\frac{l_{11}-l_{22}+i(l_{21}+l_{12})}{2}.
 \label{sm:swap:complex}
\end{equation}
Then $|a_\sigma|^2-|b_\sigma|^2=1$ and
\begin{align}
 \mathbb E b_\sigma&=0,\qquad \mathbb E a_\sigma=a_0e^{i\theta},
 \nonumber\\
 a_\sigma&=\sqrt{1+|b_\sigma|^2}\,e^{i(\theta+d_\sigma)},\qquad
 \max_\sigma(|b_\sigma|+|d_\sigma|)\le C_0|E|.
 \label{sm:swap:center}
\end{align}
Write $B^2=\mathbb E|b_\sigma|^2$ and $U^2=\mathbb E d_\sigma^2$.
The variance $B^2$ is denoted $\mathcal V(E)$ in the main text.
The coefficient $b_\sigma$ mixes a complex amplitude with its conjugate
and is the mixing term responsible for reflection in this basis. Its
mean vanishes by the choice of basis; phase estimates connect its variance
to the leading accumulated logarithmic growth.
We prove the probability-uniform estimate
\begin{equation}
 \left|\gamma_{\rm block}(E)-\frac{B^2}{2}\right|
 \le C_1|E|B^2.
 \label{sm:swap:relative}
\end{equation}

Set $q_\sigma=b_\sigma/a_\sigma$ and
$c_\sigma=\bar a_\sigma/a_\sigma$. For the projective coordinate
$z=\bar v/v$, $|z|=1$, the exact update and logarithmic stretch are
\begin{equation}
 z'=c_\sigma\frac{z+\bar q_\sigma}{1+q_\sigma z},\qquad
 \log\|L_\sigma v\|=\log|a_\sigma|+\Re\log(1+q_\sigma z).
 \label{sm:swap:phase}
\end{equation}
The imaginary part of Eq.~\eqref{sm:swap:center} gives
$|\mathbb E d_\sigma|\le C|E|(U^2+B^2)$.
For each $k=1,2$, this implies
\begin{equation}
 |1-\mathbb E c_\sigma^k|\ge c_k|E|,
 \label{sm:swap:denominator}
\end{equation}
since $\mathbb E c_\sigma^k-e^{-2ik\theta}
=O(U^2+|E|B^2)=O(E^2)$.
The exact identity $\mathbb E b_\sigma=0$ further gives
\begin{equation}
 |\mathbb E q_\sigma|+
 |\mathbb E(c_\sigma^kq_\sigma)|+
 |\mathbb E(c_\sigma^k\bar q_\sigma)|
 \le C_k(BU+|E|B^2).
 \label{sm:swap:forcing}
\end{equation}
Indeed, subtracting the constant phase from
$e^{-id_\sigma}/\sqrt{1+|b_\sigma|^2}$ bounds the phase term by
$\mathbb E|b_\sigma d_\sigma|\le BU$ and the modulus term by
$\mathbb E|b_\sigma|^3\le C|E|B^2$; the other two terms are identical
estimates with different fixed phase multiples.

Let $\nu$ be any stationary projective measure realizing the top
exponent, and put $\nu_k=\int z^k\,d\nu$. Expanding
\begin{equation}
 (z')^k=c_\sigma^k
 [z^k+k\bar q_\sigma z^{k-1}-kq_\sigma z^{k+1}]
 +O_k(|q_\sigma|^2)
\end{equation}
and using independence of the next block from the incoming phase,
Eqs.~\eqref{sm:swap:denominator}--\eqref{sm:swap:forcing} yield
\begin{equation}
 |\nu_k|\le C_k(BU+B^2)/|E|\le C_kB,\qquad k=1,2.
 \label{sm:swap:harmonics}
\end{equation}
No uniqueness of $\nu$ is assumed. Equation~\eqref{sm:swap:phase} now gives
\begin{align}
 \gamma_{\rm block}
 ={}&\tfrac12\mathbb E\log(1+|b_\sigma|^2)
 +\Re[(\mathbb E q_\sigma)\nu_1
       -\tfrac12(\mathbb E q_\sigma^2)\nu_2]
 +O(\mathbb E|q_\sigma|^3).
\end{align}
The first term is $B^2/2+O(E^2B^2)$; the two harmonics are
$O(B^2U+|E|B^3+B^3)$, and the tail is $O(|E|B^2)$.
Since $U,B\le C|E|$, this proves Eq.~\eqref{sm:swap:relative},
including $B=0$. Its constants are uniform over the probability
simplex of the fixed family.

\section{Independent local faults and a graph lower bound}
\label{sm:swap:floor}

At each complete block draw a balanced word $W$ and a boundary $J$
from a fixed finite joint law, then draw an independent
$\xi\sim\mathrm{Bernoulli}(p_{\rm sw})$. Transmit $W$ for $\xi=0$ and
$W^{(J)}$ for $\xi=1$. Successive blocks are independent.
The boundary law may depend on the word. Let $\bar A_{p_{\rm sw}},\bar B_{p_{\rm sw}}$
be the realized-word averages in Eq.~\eqref{sm:swap:derivative}.
Their elliptic normalizer at $E=0$ is
$R_{p_{\rm sw}}=\operatorname{diag}[(\bar B_{p_{\rm sw}}/\bar A_{p_{\rm sw}})^{1/4},
(\bar A_{p_{\rm sw}}/\bar B_{p_{\rm sw}})^{1/4}]$.
Equation~\eqref{sm:swap:relative} gives the exact fixed-$p_{\rm sw}$
coefficient per physical cell,
\begin{equation}
 \lim_{E\to0}\frac{\gamma(E,p_{\rm sw})}{E^2}
 =\frac1{8\ell K^2}\mathbb E\left[
 \sqrt{\frac{\bar A_{p_{\rm sw}}}{\bar B_{p_{\rm sw}}}}B_{\rm realized}
 -\sqrt{\frac{\bar B_{p_{\rm sw}}}{\bar A_{p_{\rm sw}}}}A_{\rm realized}
 \right]^2.
 \label{sm:swap:coefficient}
\end{equation}

More strongly, uniformly for $p_{\rm sw}\in[0,1]$ and sufficiently small $|E|$,
\begin{equation}
 \gamma(E,p_{\rm sw})\ge\frac{p_{\rm sw}(1-p_{\rm sw})E^2}{2\ell}
 \mathbb E_{W,J}\chi_{W_J,W_{J+1}}^2(1-C|E|).
 \label{sm:swap:uniform-floor}
\end{equation}
To prove this, use linearity of the coordinate $b(L)$ in
Eq.~\eqref{sm:swap:complex} and conditional variance to obtain
\begin{equation}
 B^2\ge p_{\rm sw}(1-p_{\rm sw})\mathbb E_{W,J}
 \left|b\bigl(R[P_{W^{(J)}}-P_W]R^{-1}\bigr)\right|^2.
\end{equation}
For every real matrix $L$, including matrices of nonzero trace,
$|b(L)|^2=|a(L)|^2-\det L\ge-\det L$.
Equation~\eqref{sm:swap:invariant} and analyticity give
$-\det(P_{W^{(J)}}-P_W)=\chi_{ab}^2E^2+O(E^3)$.
The determinant is unchanged by $R$. For every positive $\chi_{ab}$,
the error can be bounded relatively by $C|E|\chi_{ab}^2E^2$;
there are only finitely many pairs. Zero-weight pairs are simply
omitted from this lower bound. Combining these facts with
Eq.~\eqref{sm:swap:relative} proves
Eq.~\eqref{sm:swap:uniform-floor}. It therefore holds along arbitrary
probabilities $p_{\rm sw}=p_{\rm sw}(E)$, however rapidly they vanish.

Suppose now that every word contains the same positive multiplicities
of $m\ge2$ distinct hoppings, and $J$ is uniform over the $\ell-1$
internal boundaries. On the complete alphabet graph assign edge
weight $\chi_{ab}^2$, and let $W_{\rm tree}$ be its minimum
spanning-tree weight. Then
\begin{equation}
 \sum_{j=1}^{\ell-1}\chi_{w_j,w_{j+1}}^2\ge W_{\rm tree}>0,
 \qquad
 \gamma(E,p_{\rm sw})\ge
 \frac{p_{\rm sw}(1-p_{\rm sw})W_{\rm tree}}{2\ell(\ell-1)}E^2(1-C|E|).
 \label{sm:swap:graph}
\end{equation}
The adjacent-edge multigraph of each word is connected and spans the
alphabet, so it contains a spanning tree of no greater total weight.
For strict positivity put $f_a=(t_a^2-K^2)/Q_a$.
By Eq.~\eqref{sm:swap:invariant}, zero-weight edges join exactly equal
$f$ values. The balanced geometric mean $K$ lies strictly between
the extreme hoppings, so $f_{\min}<0<f_{\max}$. Every spanning tree
crosses distinct $f$ classes and has positive weight. There are
finitely many trees. In the physical affine-metric model, every
unequal edge has positive weight and
$W_{\rm tree}\ge(m-1)\min_{a\ne b}\chi_{ab}^2$.

The leading floor in Eq.~\eqref{sm:swap:graph} is independent of
the choice and ordering of block words. Its small-energy
constants can also be chosen uniformly over all distributions of words with
the fixed cell types and composition, because their set of words is
finite and Eq.~\eqref{sm:swap:derivative} gives uniform ellipticity.
Thus independent local transpositions restore a quadratic
localization floor at every fixed $0<p_{\rm sw}<1$, while composition still
preserves exact zero-energy transparency.

The statement concerns fixed-length iid blocks with one independent
swap coin per block. It does not cover growing blocks or correlated
repairs. The factor $1-p_{\rm sw}$ cannot be discarded near $p_{\rm sw}=1$: always
swapping a periodic word can produce another periodic word.
In the general positive-metric class, swaps deliberately restricted
to zero-weight edges need not give a floor. Parameter-uniform bounds
require a compact feasible alphabet and positive metrics; a map
collapsing all levels removes the disorder. At fixed $E,p_{\rm sw}$, bounded
contacts give $-\log\mathcal T_N/(2N)\to\gamma(E,p_{\rm sw})$.
Equation~\eqref{sm:swap:graph} bounds the bulk localization indicator;
finite-device resistance moments in a joint length--energy limit
require a separate finite-product analysis.

\section{A control at equal block length and entropy}
\label{sm:fault:control}

Use the reverse seeds and the finite-energy transfers of
Secs.~\ref{sm:energy}--\ref{sm:energy:hierarchy}. Fix $r\ge1$, put
$m=2^r$ and $\ell=4m$, and define the repeated-seed control
\begin{equation}
 R_+(E)=A_0(E)^m,\qquad R_-(E)=B_0(E)^m.
 \label{sm:fault:repeats}
\end{equation}
Choose these two words independently with equal probabilities.
Each control word and each ideal word $I_+=A_r$, $I_-=B_r$ has the
same length $\ell$, with $\ell/4$ copies of each extreme hopping and
$\ell/2$ copies of the middle hopping. Both ensembles have one fair bit
per block, or entropy $\log2/\ell$ per cell, and both transmit perfectly
at zero energy through any number of complete blocks. Their physical
source magnitudes are identical. Additional independent source signs
contribute the same extra entropy and site-averaged white spectrum in
both cases without changing any transfer matrix.

At the common identity resonance, differentiation of the powers gives
\begin{equation}
 R_+-R_-=mEC_0+O(E^2),\qquad
 \frac{R_+'(0)+R_-'(0)}2=m\omega_0J.
\end{equation}
The Lyapunov lemma of Sec.~\ref{sm:energy:lyapunov} yields, per cell,
\begin{equation}
 \gamma_{{\rm rep},r}(E)=\frac{m c_0^2}{32}E^2+o(E^2),
 \qquad
 \gamma_{{\rm id},r}(E)=\frac{c_r^2}{8\ell}E^{2r+2}+o(E^{2r+2}).
 \label{sm:fault:fair-comparison}
\end{equation}
The improved exponent therefore requires the ordering constraint, not
just a longer correlation block or a lower entropy rate. This comparison
does not assert optimality among all words with those resources.

\section{A joint ordering-fault localization law}
\label{sm:fault:law}

Independently at each length-$\ell$ block, use the four-word distribution
\begin{equation}
 \Pr(I_+)=\Pr(I_-)=\frac{1-p_{\rm rep}}{2},\qquad
 \Pr(R_+)=\Pr(R_-)=\frac{p_{\rm rep}}{2},
 \qquad 0\le p_{\rm rep}\le1.
 \label{sm:fault:ensemble}
\end{equation}
The fault probability $p_{\rm rep}$ is per block, not per physical cell. Every
word has the same composition, so even an arbitrary fault pattern
preserves $\mathcal T_N(0)=1$ for complete blocks under exact bulk
retuning and matched contacts.

For each fixed depth and fixed feasible strictly ordered hopping
alphabet, the Lyapunov exponent per cell obeys
\begin{equation}
 \gamma_r(E,p_{\rm rep})=\frac{(1-p_{\rm rep})c_r^2}{8\ell}E^{2r+2}
 +\frac{p_{\rm rep} m c_0^2}{32}E^2
 +o\!\left(|E|^{2r+2}+p_{\rm rep}E^2\right).
 \label{sm:fault:joint-lyapunov}
\end{equation}
The remainder divided by the displayed scale tends to zero uniformly
for $p_{\rm rep}\in[0,1]$ as $E\to0$. In the joint limit $E\to0$, $p_{\rm rep}\to0$,
the factor $1-p_{\rm rep}$ in the ideal term may consequently be replaced by one.

\subsection{Centering and uniform estimates}

Define
\begin{equation}
 \mathcal I=\frac{I_++I_-}{2},\quad
 \mathcal R=\frac{R_++R_-}{2},\quad
 C=(1-p_{\rm rep})\mathcal I+p_{\rm rep}\mathcal R,\quad
 B=\mathcal R-\mathcal I.
\end{equation}
The ideal and repeat means have the same first derivative, hence
\begin{equation}
 B=O(E^2),\qquad C=I+m\omega_0EJ+O(E^2)
 \label{sm:fault:mean}
\end{equation}
uniformly in $p_{\rm rep}$. Let $\Delta_\sigma=M_\sigma-C$ for any of the four
transfers and write
\begin{align}
 U&=(I_+-I_-)/2=E^{r+1}C_r/2+O(E^{r+2}),\nonumber\\
 V&=(R_+-R_-)/2=mEC_0/2+O(E^2),\nonumber\\
 \Delta_{I\pm}&=\pm U-p_{\rm rep}B,\qquad
 \Delta_{R\pm}=\pm V+(1-p_{\rm rep})B.
 \label{sm:fault:centered}
\end{align}
For the scale $a(E,p_{\rm rep})=|E|^{2r+2}+p_{\rm rep}E^2$, this gives
\begin{equation}
 \sum_\sigma w_\sigma\|\Delta_\sigma\|^2=O(a),\qquad
 \max_\sigma\|\Delta_\sigma\|=O(|E|).
 \label{sm:fault:weighted-estimates}
\end{equation}
The exact variance decomposition cancels all within-pair cross terms.
More explicitly,
\begin{align}
 \sum_\sigma w_\sigma\Delta_\sigma^T\Delta_\sigma
 &=(1-p_{\rm rep})U^TU+p_{\rm rep}V^TV+p_{\rm rep}(1-p_{\rm rep})B^TB,\nonumber\\
 \sum_\sigma w_\sigma\det\Delta_\sigma
 &=(1-p_{\rm rep})\det U+p_{\rm rep}\det V+p_{\rm rep}(1-p_{\rm rep})\det B.
 \label{sm:fault:variance-identities}
\end{align}
The second identity uses that the determinant is a quadratic form on
$2\times2$ matrices. Taking the trace of the first identity gives the
exact Frobenius-norm variance decomposition. Its between-group term is
$p_{\rm rep}(1-p_{\rm rep})\|B\|_F^2=O(p_{\rm rep}E^4)=o(a)$; the operator-norm estimates follow
from these Frobenius bounds.
These estimates hold irrespective of the rate at which $p_{\rm rep}$ vanishes.

Normalize $C/\sqrt{\det C}$ to a rotation with a real basis
$Q(E,p_{\rm rep})=I+O(E)$. Its angle is
$\theta=m\omega_0E+O(E^2)$, uniformly in $p_{\rm rep}$, by
Eq.~\eqref{sm:fault:mean}. Put $D_\sigma=C^{-1}\Delta_\sigma$ and let
a prime denote conjugation by $Q$. Exact centering and the determinant
conditions give
\begin{equation}
 \sum_\sigma w_\sigma D_\sigma=0,\quad
 \det C\,(1+\overline d)=1,\quad
 \operatorname{tr}D_\sigma=\overline d-\det D_\sigma,
 \qquad \overline d=\sum_\sigma w_\sigma\det D_\sigma.
 \label{sm:fault:determinant}
\end{equation}
The individual centered matrices need not be traceless; the last
identity retains their scalar parts.

\subsection{Angular and radial limits}

For a unit vector $v_0$, expand the expected logarithmic stretch in the
rotating basis. The linear noise term vanishes exactly, and
\begin{align}
 h(v_0)={}&-\frac12\overline d
 +\frac12\sum_\sigma w_\sigma\|D_\sigma'v_0\|^2
 -\sum_\sigma w_\sigma(v_0^TD_\sigma'v_0)^2
 +O(|E|a+a^2).
 \label{sm:fault:radial}
\end{align}
Indeed, the expected cubic remainder is bounded by
$\max\|D_\sigma'\|\sum w_\sigma\|D_\sigma'\|^2=O(|E|a)$;
expansion of the scalar normalization contributes $O(a^2)$.
Using Eq.~\eqref{sm:fault:centered}, the leading quadratic terms come
from $\pm E^{r+1}C_r/2$ and $\pm mEC_0/2$, with weights $1-p_{\rm rep}$ and $p_{\rm rep}$.
The omitted remainder is $o(a)$. The group-mean displacement contributes only
$O(p_{\rm rep}E^4)$.

For a smooth function $f$ of projective angle, stationarity gives
\begin{equation}
 \theta\int f'\,d\nu_{E,p_{\rm rep}}=O(E^2+a).
\end{equation}
Since $a=O(E^2)$ uniformly in $p_{\rm rep}$, division by $\theta$ makes every
weak limiting stationary measure uniform. This holds along every
sequence of probabilities $p_{\rm rep}=p_{\rm rep}(E)$, including arbitrarily rare faults.
The angular average of the quadratic expression in
Eq.~\eqref{sm:fault:radial} is $c^2/2$ for a symmetric traceless matrix
with eigenvalues $\pm c$. Applying this to $C_r/2$ and $mC_0/2$,
then dividing by $\ell$, proves Eq.~\eqref{sm:fault:joint-lyapunov}.
Uniformity follows by subsequences: the two quadratic weights divided
by $a$ are bounded, and every convergent subsequence has the same
uniform angular limit. No fixed lower bound on $p_{\rm rep}$ was used.

\section{Exact device bounds and the mean resistance crossover}
\label{sm:fault:resistance}

To connect the bulk localization law to a device, we retain the exact
one-block second moment and iterate it before taking any joint
length--energy limit.

The two-dimensional determinant identity and centering imply exactly
\begin{equation}
 \det C+\sum_\sigma w_\sigma\det\Delta_\sigma=1,\qquad
 \mathbb E(M_\sigma'^TM_\sigma')
 =\det C\,I+\sum_\sigma w_\sigma\Delta_\sigma'^T\Delta_\sigma'.
 \label{sm:fault:second-moment}
\end{equation}
Define
\begin{equation}
 q(E,p_{\rm rep})=\lambda_{\max}\!\left(
 \sum_\sigma w_\sigma\Delta_\sigma'^T\Delta_\sigma'\right)
 -\sum_\sigma w_\sigma\det\Delta_\sigma\ge0.
 \label{sm:fault:q}
\end{equation}
Its nonnegativity follows because the positive matrix on the left of
Eq.~\eqref{sm:fault:second-moment} has trace at least two. For $n$
independent blocks, conditional expectation gives
$\mathbb E\|P_n'\|_F^2\le2[1+q(E,p_{\rm rep})]^n$.

The corresponding exact lower factor is
\begin{equation}
 q_-(E,p_{\rm rep})=\lambda_{\min}\!\left(
 \sum_\sigma w_\sigma\Delta_\sigma'^T\Delta_\sigma'\right)
 -\sum_\sigma w_\sigma\det\Delta_\sigma.
 \label{sm:fault:q-lower}
\end{equation}
It need not be nonnegative. However, $1+q_->0$ because it is the
smallest eigenvalue of the positive-definite matrix
$\mathbb E(M'^TM')$. Conditioning each factor gives
$\mathbb E\|P_n'\|_F^2\ge2(1+q_-)^n$ at every energy where the
mean-rotation basis is defined.

With the finite-energy lead metric $L(E)$ of
Eq.~\eqref{sm:energy:lead-metric}, put
$k(E,p_{\rm rep})=\operatorname{cond}_2[L(E)Q(E,p_{\rm rep})^{-1}]$. The exact transmission
identity gives, for $N=\ell n$,
\begin{align}
 \mathbb E(\mathcal T_N^{-1}-1)
 &\le\frac{k(E,p_{\rm rep})^2[1+q(E,p_{\rm rep})]^n-1}{2},
 \label{sm:fault:inverse-bound}\\
 \Pr\{\mathcal T_N<1-\varepsilon\}
 &\le\frac{1-\varepsilon}{2\varepsilon}
 \left[k(E,p_{\rm rep})^2[1+q(E,p_{\rm rep})]^n-1\right],\qquad0<\varepsilon<1.
 \label{sm:fault:probability}
\end{align}
The matching computable resistance lower bound is
\begin{equation}
 \mathbb E(\mathcal T_N^{-1}-1)\ge
 \max\left\{0,\frac{k(E,p_{\rm rep})^{-2}[1+q_-(E,p_{\rm rep})]^n-1}{2}\right\}.
 \label{sm:fault:exact-lower}
\end{equation}
This finite-$N$ lower certificate does not require a Taylor remainder
constant. Its truncation at zero uses the physical inequality
$\mathcal T_N\le1$; it does not replace $q_-$ by zero before exponentiation.
Here $q(E,p_{\rm rep})=O(a)$ and $k(E,p_{\rm rep})=1+O(|E|)$ uniformly in $p_{\rm rep}$. Thus
$n(|E|^{2r+2}+p_{\rm rep}E^2)$ is the relevant finite-device scale.

\subsection{A matching second-moment lower bound}

Set
\begin{equation}
 v(E,p_{\rm rep})=\frac12\left[(1-p_{\rm rep})c_r^2E^{2r+2}
                    +p_{\rm rep} m^2c_0^2E^2\right].
 \label{sm:fault:scale}
\end{equation}
The leading centered matrices in Eq.~\eqref{sm:fault:centered} are
symmetric and traceless. Their squares are scalar, and their
determinants are the negatives of those scalars. Consequently
\begin{equation}
 \mathbb E(M'^TM')=[1+v(E,p_{\rm rep})]I+\mathcal R(E,p_{\rm rep}),\qquad
 \|\mathcal R(E,p_{\rm rep})\|\le C|E|a(E,p_{\rm rep}).
 \label{sm:fault:isotropic}
\end{equation}
The scale $v$ is uniformly comparable to $a$ for small $E$ and
$p_{\rm rep}\in[0,1]$. For $p_{\rm rep}\le1/2$ the ideal term retains at least half its
weight; for $p_{\rm rep}>1/2$ the fault term dominates the possible missing ideal
contribution. Iterating the two operator inequalities in
Eq.~\eqref{sm:fault:isotropic} yields
\begin{equation}
 2[1+v-C|E|a]^n\le\mathbb E\|P_n'\|_F^2
 \le2[1+v+C|E|a]^n.
 \label{sm:fault:sandwich}
\end{equation}
The lower factor is positive for sufficiently small $|E|$.
Changing to the lead basis multiplies these bounds by $k^{-2}$ and
$k^2$, respectively.

Consider arbitrary sequences $E\to0$, $p_{\rm rep}\in[0,1]$, and integer
$n\to\infty$. If $n v(E,p_{\rm rep})\to s\in[0,\infty)$, then
$n|E|a\to0$ and $nv^2\to0$. Since $k\to1$, the exact norm identity
for transmission gives
\begin{equation}
 \boxed{\mathbb E(\mathcal T_N^{-1}-1)
       \longrightarrow\frac{e^s-1}{2}.}
 \label{sm:fault:mean-limit}
\end{equation}
If instead $nv\to\infty$, Eq.~\eqref{sm:fault:sandwich} implies
\begin{equation}
 \frac{\log\mathbb E(\mathcal T_N^{-1})}{n v(E,p_{\rm rep})}\longrightarrow1.
 \label{sm:fault:divergence}
\end{equation}
These are limits of the mean inverse transmission, or mean excess
Landauer resistance in its natural resistance unit. An arithmetic mean
conductance or a median cannot be obtained by inverting that mean.
The proof uses the finite-length moment recursion directly, rather than
interchanging a fixed-energy infinite-length Lyapunov limit with a
finite-size limit. The two independently derived quantities satisfy
$\ell\gamma_r(E,p_{\rm rep})=v(E,p_{\rm rep})/4+o(a)$.

\section{Fault budget at the ideal energy scale}
\label{sm:fault:budget}

At a fixed-depth energy $E=bN^{-1/(2r+2)}$ with $b>0$ fixed,
\begin{equation}
 n v(E,p_{\rm rep})=\frac{(1-p_{\rm rep})c_r^2b^{2r+2}}{2\ell}
       +\frac{m^2c_0^2b^2}{2\ell}\,p_{\rm rep}N^{r/(r+1)}.
 \label{sm:fault:scaled-budget}
\end{equation}
It follows that
\begin{equation}
 p_{\rm rep}N^{r/(r+1)}=O(1)
 \label{sm:fault:budget-formula}
\end{equation}
is sufficient for bounded mean excess resistance, and necessary for
that observable in the four-block fault ensemble. If the quantity in
Eq.~\eqref{sm:fault:budget-formula} is unbounded, choose a subsequence
along which it diverges; Eq.~\eqref{sm:fault:divergence} makes the mean
inverse transmission unbounded on that subsequence. A sufficiently
small budget constant and energy prefactor also give any specified
high-transmission probability through Eq.~\eqref{sm:fault:probability}.
This does not assert an optimal fault budget over other rules for assembling the chain.

For fixed $p_{\rm rep}>0$, the asymptotic exponent is quadratic. Comparing the
two terms of Eq.~\eqref{sm:fault:joint-lyapunov} gives the crossover
\begin{equation}
 |E_\times|^{2r}\asymp
 \frac{p_{\rm rep} m^2c_0^2}{(1-p_{\rm rep})c_r^2},\qquad0<p_{\rm rep}<1.
\end{equation}
Ordering faults leave the central resonance intact but change the
surrounding scattering law. They differ from a residual mean error
$K_{\rm used}=K_Fe^\eta$, for which
$\mathcal T_N(0)=\operatorname{sech}^2(N\eta)$ when contacts are matched
to $K_{\rm used}$. The latter still requires inverse-length accuracy.

All estimates keep the depth fixed. They are uniform on compact feasible
strictly ordered calibration sets with positive hopping lower bounds
and bounded finite-mediator parameters. The random choices in
Eq.~\eqref{sm:fault:ensemble} are independent between blocks, and the
near-unit probability claims assume matched contacts. The interval
of energies is understood pointwise as in
Sec.~\ref{sm:energy:certificate}; a joint guarantee over a continuum of
energies in one sample is not inferred.

\section{The full resistance distribution}
\label{sm:diffusion}

The limiting one-channel Melnikov/DMPK diffusion is established transport
theory \cite{Beenakker1997,Bachmann2012}. Here we derive its microscopic
length scale and verify its applicability at the identity resonance,
including arbitrarily rare ordering faults. Fix a finite depth $r\ge1$
and the feasible strictly ordered hopping alphabet of
Sec.~\ref{sm:fault:law}. Keep exact mean retuning and matched finite-energy
contacts. For any sequence $E\to0$, $E\ne0$, and $p_{\rm rep}=p_{\rm rep}(E)\in[0,1]$, let
$v=v(E,p_{\rm rep})$ be Eq.~\eqref{sm:fault:scale}. The accumulated
second-moment growth is measured by $nv$, so a nontrivial resistance
distribution requires lengths of order $1/v$ as scattering weakens.
Define
\begin{equation}
 n_E(s)=\lfloor s/v\rfloor,\qquad
 \rho_E(s)=\mathcal T_{\ell n_E(s)}(E)^{-1}-1.
 \label{sm:diffusion:process}
\end{equation}
On each bounded $s$ interval this process converges in distribution in
the Skorokhod topology to the continuous diffusion
\begin{equation}
 d\rho_s=(\tfrac12+\rho_s)\,ds
        +\sqrt{\rho_s(1+\rho_s)}\,dW_s,\qquad \rho_0=0.
 \label{sm:diffusion:radial-sde}
\end{equation}
Its generator and forward equation are
\begin{equation}
 \mathcal Lf=\tfrac12\rho(1+\rho)f''+(\tfrac12+\rho)f',\qquad
 \partial_sP=\tfrac12\partial_\rho[\rho(1+\rho)\partial_\rho P],
 \quad P(\rho,0)=\delta_0.
 \label{sm:diffusion:generator}
\end{equation}
The length normalization is fixed by the microscopic $v$, rather than
by an independently fitted mean free path.

\subsection{Exact mean removal and increment estimates}

Retain the four transfers $M_\sigma$, their weights $w_\sigma$, and the
centered matrices $C=\sum w_\sigma M_\sigma$,
$\Delta_\sigma=M_\sigma-C$ from Sec.~\ref{sm:fault:law}. The basis
$Q(E,p_{\rm rep})=I+O(E)$ makes
\begin{equation}
 C'=\alpha R_\theta,\qquad \alpha=\sqrt{\det C},\qquad
 \theta=m\omega_0E+O(E^2),\qquad
 M_\sigma'=R_\theta(\alpha I+\xi_\sigma),\quad
 \xi_\sigma=R_\theta^{-1}\Delta_\sigma'.
 \label{sm:diffusion:centering}
\end{equation}
A prime denotes conjugation by $Q$, and $\sum w_\sigma\xi_\sigma=0$
exactly. All estimates below are uniform in $p_{\rm rep}\in[0,1]$.
The leading centered symmetric matrices $S_\sigma$ are
$\pm E^{r+1}C_r/2$ in the ideal pair and $\pm mEC_0/2$ in the repeat
pair. Equation~\eqref{sm:fault:centered} gives
\begin{align}
 \sum_\sigma w_\sigma\|\xi_\sigma-S_\sigma\|^2&=O(E^2v),&
 \sum_\sigma w_\sigma\|S_\sigma\|^2&=O(v),\nonumber\\
 \max_\sigma\|\xi_\sigma\|&=O(|E|),&
 \sum_\sigma w_\sigma\|\xi_\sigma\|^3&=O(|E|v),\nonumber\\
 \alpha&=1+v/4+O(|E|v).
 \label{sm:diffusion:estimates}
\end{align}
The covariance error is therefore $O(|E|v)$. In deriving these bounds,
the ideal and repeat means differ by $O(E^2)$ and contribute the
weighted variance $O(p_{\rm rep}E^4)=O(E^2v)$. Their mean displacement is included
in $C$, so it generates no omitted drift. Finally,
$\sum w_\sigma\det\Delta_\sigma=-v/2+O(|E|v)$ and
Eq.~\eqref{sm:fault:second-moment} prove the expansion of $\alpha$.

For $P_k'=M_k'\cdots M_1'$, remove the mean rotation exactly:
\begin{equation}
 Y_k=R_\theta^{-k}P_k',\qquad
 Y_{k+1}=(\alpha I+\zeta_{k+1})Y_k,\quad
 \zeta_{k+1}=R_\theta^{-k}\xi_{\sigma_{k+1}}R_\theta^k,\quad Y_0=I.
 \label{sm:diffusion:recursion}
\end{equation}
The increments $\zeta_k$ are independent and centered, with
deterministically varying covariances. For
$X=\left(\begin{smallmatrix}0&1\\1&0\end{smallmatrix}\right)$ and
$Z=\left(\begin{smallmatrix}1&0\\0&-1\end{smallmatrix}\right)$,
conjugation rotates their coefficient plane by $2k\theta$. Thus the
entrywise covariance has the form
\begin{equation}
 \mathbb E(\zeta_{k+1}\otimes\zeta_{k+1})
 =\frac v4(X\otimes X+Z\otimes Z)
 +v\{\mathsf A\cos(4k\theta)+\mathsf B\sin(4k\theta)\}
 +O(|E|v),
 \label{sm:diffusion:covariance}
\end{equation}
with uniformly bounded tensors $\mathsf A,\mathsf B$. They need not
converge as the relative ideal and fault weights change. Crucially,
$v/|\theta|=O(|E|)\to0$.
This scale separation makes the rotation explore many angles before
scattering produces an order-one change, which is why angular averaging
is needed below.

\subsection{Averaging with the evolving state}

The oscillatory covariance must also average when tested against the
random state. Stop $Y_k$ on leaving a fixed compact set. For a bounded
smooth $F$, its one-step conditional drift and second moment are both
$O(v)$ by Eqs.~\eqref{sm:diffusion:estimates} and
\eqref{sm:diffusion:recursion}. Discrete summation by parts gives
\begin{align}
 \sum_{k<n}v e^{4ik\theta}F(Y_k)
 =\frac{v}{1-e^{4i\theta}}\biggl[
 F(Y_0)-e^{4in\theta}F(Y_n)
 +\sum_{k<n}e^{4i(k+1)\theta}
       \{F(Y_{k+1})-F(Y_k)\}\biggr].
 \label{sm:diffusion:abel}
\end{align}
For $n\le S/v$, the bracketed increment sum has drift $O(S)$ and a
martingale part $O_P(\sqrt S)$, uniformly up to that time by the
martingale maximal inequality. Its boundary terms are bounded; a
stopping indicator adds at most one bounded jump. Hence the left side
of Eq.~\eqref{sm:diffusion:abel} is
$O_P[(v/|\theta|)(1+S+\sqrt S)]$ and tends to zero. This proves
state-dependent covariance averaging without a random-phase assumption.

\subsection{The matrix martingale problem}

For a compactly supported $C^4$ test function $f$, the expected cubic
Taylor remainder in Eq.~\eqref{sm:diffusion:recursion} is $O(|E|v)$ on
compact sets. Its sum over $O(1/v)$ steps vanishes. The scalar mean
contributes $(v/4)Df(Y)[Y]$. Covariance averaging leaves
$(v/8)\sum_{A=X,Z}D^2f(Y)[AY,AY]$. Every subsequential limit therefore
has generator
\begin{equation}
 \mathcal Gf(Y)=\frac14Df(Y)[Y]
              +\frac18\sum_{A=X,Z}D^2f(Y)[AY,AY].
 \label{sm:diffusion:matrix-generator}
\end{equation}
To verify tightness, the drift of the recursion is bounded by
$Cv\|Y\|$ and its conditional martingale variance by $Cv\|Y\|^2$.
The martingale maximal inequality and discrete Gronwall give
$\mathbb E\sup_{k\le S/v}\|Y_k\|^2\le C_S$. This gives compact
containment. After compact stopping, increments over a rescaled time
$\delta$ have second moment $O(\delta+v)$, and the maximum single jump
vanishes by $\max\|\xi_\sigma\|=O(E)$. The processes are consequently
tight with continuous limiting paths, and the stopping can be removed.

The martingale problem has the unique nonexplosive solution
\begin{equation}
 dY_s=\frac14Y_s\,ds
      +\frac12(X\,dW_s^{(1)}+Z\,dW_s^{(2)})Y_s,\qquad Y_0=I,
 \label{sm:diffusion:matrix-sde}
\end{equation}
where the Brownian motions are independent. Its coefficients are
globally Lipschitz in the matrix entries. In Stratonovich form its
drift is zero because $X^2=Z^2=I$, so $\det Y_s=1$, as in every
microscopic product. This uniqueness identifies the full limit.
Independence and exact centering of the increments preclude an
additional area drift; the second-order Taylor terms are all retained
in Eq.~\eqref{sm:diffusion:matrix-generator}.

\subsection{Radial projection, contacts, and moments}

Put $\widehat\rho(Y)=(\|Y\|_F^2-2)/4$. Writing
$YY^T=(1+2\rho)I+xX+zZ$ gives
$x^2+z^2=4\rho(1+\rho)$ by unit determinant. For the left-invariant
derivatives $V_A f(Y)=\partial_t f(e^{tA}Y)|_{t=0}$,
\begin{equation}
 V_X\rho=x,\quad V_Z\rho=z,\qquad
 V_X^2\rho=V_Z^2\rho=2+4\rho.
\end{equation}
Since $\mathcal G=(V_X^2+V_Z^2)/8$, its action on $f(\rho)$ is
Eq.~\eqref{sm:diffusion:generator}. The radial diffusion has a unique
nonnegative law, with existence also supplied by the matrix
representation. Starting from zero it enters the positive half-line.

For the physical transmission, write $G=L(E)Q(E,p_{\rm rep})^{-1}=I+O(E)$.
The exact scattering identity yields
\begin{equation}
 \rho_E(s)=\frac{\|G R_\theta^{n_E(s)}Y_{n_E(s)}G^{-1}\|_F^2-2}{4}.
\end{equation}
The rotation leaves a Frobenius norm unchanged. On compact matrix sets
the difference from $\widehat\rho(Y_{n_E(s)})$ is uniformly $O(|E|)$,
independently of the fast angle. Compact containment then gives the
same limit for the actual decorated-chain transmission. Fixed unmatched
contacts need not have $G\to I$ and are not included in this statement.

The radial generator gives
\begin{align}
 \mathbb E\rho_s&=(e^s-1)/2,\nonumber\\
 \mathbb E\rho_s^2&=e^{3s}/6-e^s/2+1/3,\nonumber\\
 \operatorname{Var}\rho_s&=(2e^{3s}-3e^{2s}+1)/12.
 \label{sm:diffusion:moments}
\end{align}
These are also limits of the microscopic moments. For every fixed
integer $q\ge1$, expand the homogeneous polynomial $\|Y\|_F^{2q}$
in Eq.~\eqref{sm:diffusion:recursion}. The centered linear term vanishes,
and higher terms are bounded by the second moment times bounded
increment factors. Thus
$\mathbb E[\|Y_{k+1}\|_F^{2q}\mid Y_k]
\le(1+C_qv)\|Y_k\|_F^{2q}$, giving a uniform bound up to $S/v$.
Taking $q>4$ proves uniform integrability of $\rho_E$ and $\rho_E^2$,
and establishes Eq.~\eqref{sm:diffusion:moments} beyond weak convergence.

Rare faults satisfy the same limit because each fault scattering
increment is $O(E)\to0$. If faults contribute a finite fraction of $v$,
their count on the scale $n\sim1/v$ is of order $E^{-2}$ and diverges.
If only finitely many remain, their scattering contribution vanishes.
The weighted third-moment bound gives the corresponding Lindeberg
condition directly. This differs from finite-amplitude zero-energy
logarithmic-mass jumps, which can have a compound-Poisson limit.

The convergence describes the length process at each specified energy,
with $E\to0$ between members of the sequence. It is not a joint
distribution across several energies in one device, and does not
establish a simultaneous continuum-energy guarantee. Depth stays fixed,
block choices are independent, mean retuning is exact, and contact
matching is imposed. The limiting DMPK distribution is not a new
universality class; the calibrated microscopic scale $v(E,p_{\rm rep})$ and the
verified averaging and rare-fault conditions connect it to this model.

\section{Numerical methods for finite devices}
\label{sm:ordering:numerics}

The examples have $h=2$, $g=1.2$, $t_{\rm off}=0.25$, and the initial
hopping levels $(e^{-1},1,e)$. An additive offset $\delta$ changes $a=1.69+\delta$
and retains $u_q^2=e^q-0.25$. The intercell, lead, and contact hoppings
are all $K_\delta=[(e^{-1}+\delta)(1+\delta)^2(e+\delta)]^{1/4}$.
For the separate saturating example, $F(t)=t/(1+t/2)$, $a=1.69$,
and $u_q^2=F(e^q)-0.25>0$; these source magnitudes are re-embedded
after the map. This example tests a nonaffine common map rather than a
fixed-source additive drift.

\subsection{Equal-resource comparison}
At depth $r$, the recursive and repeated-seed arrangements each choose
one of two length-$4\,2^r$ words with a fair bit. Both have hopping
fractions $(1/4,1/2,1/4)$ and entropy $\log2/(4\,2^r)$ per cell.
The same bit strings are used for paired comparisons across arrangements,
energies, and mean errors. There are 12 offset/length/depth groups,
each with 256 independent strings, for 3072 independently drawn strings.
Their reuse produces 188416 observations across 736 conditions; observations
at different energies on the same string are not independent replicates.

For the matched-mean scan, offsets are $0,0.2,1$, lengths are 4096
and 16384 cells, and depths are one and two. Figure 2 shows medians
and 10th--90th realization percentiles, not confidence intervals for
the ensemble mean. At $N=16384$, $E=0.002$, and zero offset, the
depth-one recursive/repeated medians are $0.972523$ and
$1.26337\times10^{-7}$; the depth-two values are $0.999480$ and
$5.26343\times10^{-15}$. Stabilized matrix products retain the
logarithmic normalization, including extremely small transmissions.

The additional mean-error scan uses $K_{\rm act}=K_\delta e^\eta$
with leads rematched to $K_{\rm act}$. Its zero-energy identity is
$\mathcal T(0)=\operatorname{sech}^2(N\eta)$. A joint grid of nine
energies from zero to $0.002$, including five near-zero values between
$5\times10^{-6}$ and $10^{-4}$, and five total masses
$N\eta\in\{-0.3,-0.15,0,0.15,0.3\}$ contains 45 probes.
For depth two at zero offset, all 256 recursive samples exceed 0.9
at every grid point, whereas no repeated sample does. This describes
that finite grid; it is not a simultaneous continuum probability bound.
The exact mean-tuning tolerance for $\mathcal T(0)\ge0.9$ remains
$|\eta|\le0.3274501502\ldots/N$.

Thirty-six independent dense Green-function calculations of the full
$6N$-orbital Hamiltonian, including a nonzero mean error, check the
transfer and contact conventions with maximum absolute transmission
difference $3.22\times10^{-15}$. Formal Taylor coefficients independently
check the recursive and repeated difference orders at three offsets
and two depths. The raw observations are stored as deterministic gzip
with both compressed and decompressed SHA256 hashes, alongside sequence
seeds and hashes of the full choice arrays.

\subsection{Fault ensemble: exact moments and asymptotic comparison}
For any finite set of block matrices, the covariance obeys the exact
four-dimensional linear recursion
\begin{equation}
 \operatorname{vec}S_{n+1}
 =\left[\sum_\sigma w_\sigma
   (\widehat M_\sigma\otimes\widehat M_\sigma)\right]
 \operatorname{vec}S_n,\qquad S_0=I,
\end{equation}
where $\widehat M=L M L^{-1}$ uses the physical flux basis and
row-major vectorization. Then
$\mathbb E\rho_N=(\operatorname{tr}S_n-2)/4$.
This evaluates an ensemble moment without Monte Carlo sampling. Large
integer powers require logarithmically many matrix multiplications;
the largest plotted lengths do not represent explicit large-Hamiltonian
diagonalizations.

The fault study uses 65-decimal arithmetic for 180 scaling points,
42 fixed-device points, and 54 probability-bound probes. It covers
identity, unit-offset, and saturating calibrations and depths one and
two. To resolve the joint limit, the ideal part is set by
$n c_r^2 E^{2r+2}/2=0.1$ and
$p_{\rm rep}=\lambda c_r^2E^{2r}/(2^{2r}c_0^2)$, with
$\lambda\in\{0,0.1,1,5,20\}$ and
$N=2^{12},2^{16},2^{24},2^{32},2^{40},2^{48}$.
The actual scale retains the factor $1-p_{\rm rep}$ in its ideal contribution.
All exact finite moments lie between the two-sided bounds in the
preceding proof. At the longest length the largest relative error
against $(e^s-1)/2$ among these points is $3.43\times10^{-5}$.
This convergence check supports the normalization; the asymptotic
claim is proved analytically.

A separate complete-Hamiltonian implementation checks 27 short devices
across the three calibrations at energies $0,0.002,0.02$, with maximum
absolute transmission difference below $10^{-15}$. Enumerating all
64 three-block sequences checks the second-moment propagation directly.

An additional 55-decimal implementation constructs the $4\times4$ and
$16\times16$ tensor maps independently and contracts them for the first
and second resistance moments. All 64 three-block sequences reproduce
both contractions within $1.11\times10^{-56}$. On the specified
$p_{\rm rep}(E)\to0$ path with $s$ approaching $0.5$, increasing the block count
from $2^{20}$ to $2^{28}$ reduces the second-moment relative discrepancy
from $0.1274\%$ to $0.00797\%$ at depth one, and from $3.452\%$ to
$0.5426\%$ at depth two. The microscopic transfer formulas are shared;
the moment maps and enumeration are independent. These finite-point
checks test the diffusion normalization and its moment predictions,
while the preceding proof establishes the distributional limit.

\subsection{Outward-rounded probability enclosures}
The 63 ideal-ordering probe points shown in Fig. 2 are enclosed
independently with 60-decimal interval arithmetic. All upper bounds for
$\Pr(\mathcal T<0.9)$ are below $0.096715$. For the four-word fault
mixture, an independent 70-decimal interval implementation reconstructs
the physical cells, weighted mean, variance, invariant metric, and lead
basis. It encloses 54 probes with $N=16384$, both depths, all three
calibrations, $E\in\{10^{-4},3\times10^{-4},5\times10^{-4}\}$,
and $p_{\rm rep}\in\{10^{-6},10^{-5},10^{-4}\}$. Their largest probability
upper bound is $0.006619<0.1$.

Decimal probe values are entered as exact rationals; exponentials and
square roots are enclosed. For a positive determinant-one $2\times2$
metric, $G^{1/2}=(G+I)/\sqrt{\operatorname{tr}G+2}$ avoids an eigenvector
convention. The largest eigenvalue of a symmetric $2\times2$ matrix is
enclosed by its quadratic formula. These are pointwise energy certificates
with the stated independent-block law and matched contacts. Neither
connecting plot lines nor a Monte Carlo grid extends them to simultaneous
transparency over a continuum.

\subsection{Independent local-swap checks}
An independent first-order matrix-polynomial implementation checks the
swap determinant before using its closed formula. Across eight rational
sets of physical hopping levels and all balanced words of lengths four and eight, it
checks 17,040 unequal adjacent exchanges, 3,320 nonorthogonal conjugations,
and 64 mean-normalized conditional-variance mixtures with exact arithmetic.
An additional 80-decimal check uses the realized levels $e^{-1},1,e$.
A four-level positive-metric example also permits exchanges with zero
first-order change: $t=(1/4,1/2,2,4)$,
$Q=(15/8,3/2,6,30)$, and $K=1$. Its minimum spanning-tree weight is
81; all 24 word permutations satisfy the graph lower bound.

Figure 1(b) uses the separate local-exchange ensemble with at most one
attempt per block, uniformly over all internal boundaries including
equal-level no-ops. Exact first-derivative moments at depths one and two
give the displayed quadratic coefficients; they do not come from fitting
finite-energy Lyapunov exponents. The supplemental whole-word crossover figure uses the replacement
ensemble described above. The checks establish finite
algebraic consistency, while the preceding relative-error proof controls
the small-energy and rare-fault limits.

\paragraph{Reproducibility.}
The accompanying repository provides scripts for resource comparisons,
fault moments and interval bounds, machine-readable outputs,
compressed observations, source hashes, and figure-source records.

\subsection{Finite devices with local ordering errors}
\label{sec:local-device-numerics}

The finite-device calculation uses the same local adjacent-swap law as the
scattering obstruction.  We fix $h=2$, $g=1.2$, $t_{\rm off}=1/4$,
$t\in\{e^{-1},1,e\}$, and $K=1$, with mediator amplitudes
$u_i^2=t_i-1/4$ and core hopping $a=t_{\rm off}+2g^2/h=1.69$.
Both lead and contact hoppings equal $K$.  Each device contains
$N=16384$ cells, arranged in $1024$ independent blocks of depth $r=2$
and length $\ell=16$.  A block first chooses $A_2$ or $B_2$ with equal
probability.  With probability $p_{\rm sw}$, one of its 15 internal
boundaries is chosen uniformly and the adjacent levels are exchanged.
Equal-level choices leave the block unchanged; $p_{\rm sw}$ therefore
counts attempts, not only exchanges of unequal levels.  Every realized
block remains balanced, and every complete sample has
$\mathcal T(0)=1$.

We generate 256 independent device sequences with NumPy's PCG64 generator
and integer seed $202609180507$.  The complete arrays of base-word bits,
boundary choices, and independent uniform fault coins are reused across
all energies and all four probabilities
$p_{\rm sw}=0,0.001,0.01,0.1$.  A probability changes a coin threshold,
not the underlying block choices.  The energy grid contains 61
equally spaced points from $-0.006$ to $0.006$, including zero and
$0.002$.  Thus the independent sampling unit is one of the 256 sequences;
the $62464$ stored sample--condition observations are not independent
replicates.  Normalized double-precision $2\times2$ products evaluate
the exact finite-mediator transfer matrix.  We accumulate their logarithmic
normalizations to avoid overflow and use the matched-lead flux metric
to obtain each sample's transmission.

The solid curves in the finite-device panel are arithmetic means of the
actual transmission $\mathcal T$, not reciprocals of averaged resistance.
The shaded regions are the 10th--90th percentiles across the 256
realizations, with linear interpolation between order statistics.
They describe realization variability and are not confidence intervals
for the mean.  The figure displays $p_{\rm sw}=0,0.01,0.1$; the additional
$0.001$ condition remains in the source data.  No smoothing or fit is
applied to the sampled curves.

Independently, a 65-decimal-digit calculation propagates the covariance
by the exact finite-ensemble $4\times4$ map.  It includes both ideal
words and all 30 labeled swap choices, including no-ops.  The map yields
$\mathbb E\rho$, where $\rho=\mathcal T^{-1}-1$.  At $E=0.002$,
the results are
\begin{center}
\begin{tabular}{rccc}
\hline
$p_{\rm sw}$ & $\mathbb E\rho$ & $(1+\mathbb E\rho)^{-1}$
 & Sample mean $\mathcal T$ \\
\hline
$0$     & $0.000791937012$ & $0.999208690$ & $0.999224$ \\
$0.001$ & $0.001695337963$ & $0.998307531$ & $0.998378$ \\
$0.01$  & $0.009899563015$ & $0.990197478$ & $0.990261$ \\
$0.1$   & $0.099686414761$ & $0.909350144$ & $0.922962$ \\
\hline
\end{tabular}
\end{center}
Convexity gives the exact Jensen inequality
\begin{equation}
 \mathbb E\mathcal T
 =\mathbb E\frac{1}{1+\rho}
 \ge\frac{1}{1+\mathbb E\rho}.
\end{equation}
The third column is consequently a lower bound, not the mean transmission
or a sampled estimate of it.  The moment calculation is repeated at
$E=0,0.001,0.002,0.004,0.006$ for all four probabilities.  These are
high-precision evaluations of finite-ensemble formulas, not
outward-rounded interval certificates.  In particular, the near-unit
transmission of the finite device at a one-percent attempt rate is
compatible with a positive asymptotic quadratic localization coefficient.

Fifteen direct Green-function calculations for the full six-orbital
Hamiltonian, using five realized block words and three energies, agree
with the transfer calculation to a maximum absolute error of
$5.33\times10^{-15}$.  An exhaustive sum over the $32^2=1024$ labeled
two-block choices agrees with the covariance map within
$4.92\times10^{-59}$.  Three complete $N=16384$ sample products at
$E=0.002$ were also evaluated at 65-digit precision; their transmission
differs from the double-precision calculation by at most
$3.75\times10^{-13}$.  All sampled energy-reflection pairs satisfy the
chiral symmetry check, and all zero-energy products satisfy the balanced
center-transmission check.  The implementation guards values above one
with an explicit $2\times10^{-9}$ tolerance before any clipping;
no clipping occurred in this run.  The largest excess above unity and the unclipped
center errors are recorded with the numerical results.

The accompanying data include all $62464$ sample records, random-sequence
seeds, numerical precision checks, and software versions. The two depths
in the coefficient panel have different block lengths, so equal attempt
probabilities per block correspond to different error densities per cell.

\begin{figure}[htbp]
 \centering
 \includegraphics[width=\textwidth]{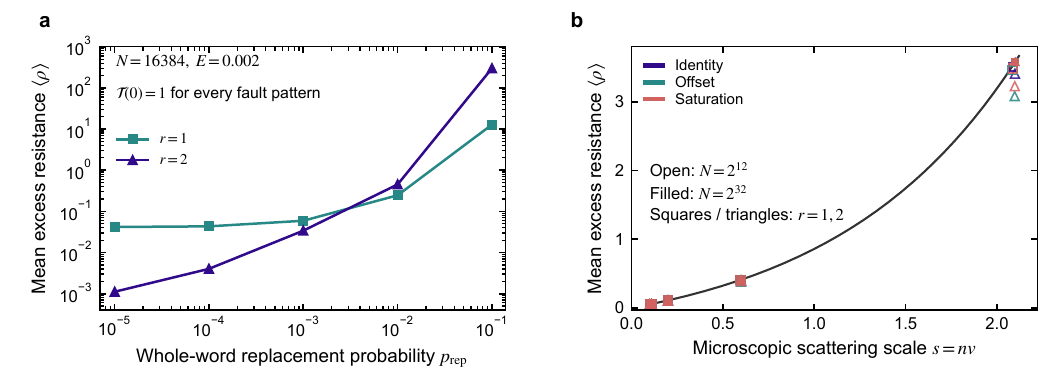}
 \caption{The separate four-word replacement ensemble.
 Left: mean excess resistance against the replacement probability
 $p_{\rm rep}$ at $N=16384$ and $E=0.002$. Right: the same mean against
 the microscopic length scale $s=nv$ for identity, offset, and saturating
 calibrations, compared with $(e^s-1)/2$. Values follow deterministic
 high-precision $4\times4$ moment propagation. These curves concern whole-word
 replacement, not the adjacent-exchange ensemble of main-text Fig.~1.}
 \label{sm:replacement:figure}
\end{figure}